\documentclass[10pt, a4paper]{article}

\usepackage{lrec-coling2024} 
\usepackage{xspace}
\usepackage{cleveref}
\usepackage{tabularx}
\usepackage{booktabs}
\usepackage{colortbl} 
\usepackage{float}
\usepackage{subfig}
\usepackage{multirow}
\usepackage{adjustbox}
\usepackage{comment}
\usepackage{hyperref}

\definecolor{green}{RGB}{183,235,198}
\definecolor{red}{RGB}{236,189,189}

\newcommand{\magpie}{MAGPIE\xspace}

\newcommand{\mtlfull}{Multi-Task Learning\xspace}
\newcommand{\mtl}{MTL\xspace}

\newcommand{\lbmfull}{Large Bias Mixture\xspace}
\newcommand{\lbm}{LBM\xspace}

\newcommand{\sotafull}{state-of-the-art\xspace}
\newcommand{\sota}{SOTA\xspace}
\title{\textbf{MAGPIE: Multi-Task Analysis of Media-Bias Generalization\\with Pre-Trained Identification of Expressions}}

\name{Tomáš Horych$^{1,3,6}$, Martin Wessel$^{4,6}$, Jan Philip Wahle$^{2}$, Terry Ruas$^{2}$,\\
{\bf \large  Jerome Waßmuth$^{5}$, André Greiner-Petter$^{2,3}$, Akiko Aizawa$^{3}$},\\
{\bf \large   Bela Gipp$^{2}$, Timo Spinde$^{2,6}$}}

\address{$^{1}$Czech Technical University, Prague, Czech Republic\\
$^{2}$University of Göttingen, Göttingen, Germany, \{\textit{last}\}@uni-goettingen.de\\
$^{3}$National Institute of Informatics, Tokyo, Japan, \{\textit{last}\}@nii.ac.jp\\
$^{4}$CDTM, TU Munich, Germany, $^{5}$University of Konstanz, Germany, \{\textit{last}\}@uni-konstanz.de\\
 $^{6}$\{t.horych, m.wessel, t.spinde\}@media-bias-research.org
 }
\abstract{
Media bias detection poses a complex, multifaceted problem traditionally tackled using single-task models and small in-domain datasets, consequently lacking generalizability. 
To address this, we introduce MAGPIE, a large-scale multi-task pre-training approach explicitly tailored for media bias detection. 
To enable large-scale pre-training, we construct \lbmfull (\lbm), a compilation of 59 bias-related tasks.
MAGPIE outperforms previous approaches in media bias detection on the Bias Annotation By Experts (BABE) dataset, with a relative improvement of 3.3\% F1-score. 
Furthermore, using a RoBERTa encoder, we show that MAGPIE needs only 15\% of fine-tuning steps compared to single-task approaches. 
We provide insight into task learning interference and show that sentiment analysis and emotion detection help learning of all other tasks, and scaling the number of tasks leads to the best results.
\magpie confirms that \mtl is a promising approach for addressing media bias detection, enhancing the accuracy and efficiency of existing models. Furthermore, \lbm is the first available resource collection focused on media bias \mtl.
 \\ \newline  }

\begin{document}

\maketitleabstract

\section{Introduction}
Media bias is a skewed portrayal of information favoring certain group interests \cite{recasens_linguistic_2013}, which manifests in multiple facets, including political, gender, racial, and linguistic biases. Such subtypes of bias, which can intersect and coexist in complex combinations, make the classification of media bias a challenging task \cite{raza_dbias_2022}. Existing research on media bias detection primarily involves training classifiers on small in-domain datasets \cite{krieger_domain-adaptive_2022}, which exhibit limited generalizability across diverse domains \cite{WesselMBIB22}.

This paper builds upon the work of \citet{WesselMBIB22}, emphasizing that the multifaceted nature of media bias detection requires a shift from isolated approaches to multi-task methodologies, considering a broad spectrum of bias types and datasets. The recent advancements in \mtlfull (\mtl) \cite{aribandi-etal-2021-reliable, chen_multi-task_2021, kirstein-etal-2022-analyzing} open up promising opportunities to overcome these challenges by enabling knowledge transfer across domains and tasks. Despite the potential, a comprehensive \mtl approach for media bias detection is yet to be realized. The only other media bias \mtl method \cite{SpindeJCDL2022} underperforms due to its narrow task focus and does not surpass baseline outcomes.

\begin{figure}[t]
    \centering
    \includegraphics[width = 0.50\textwidth]{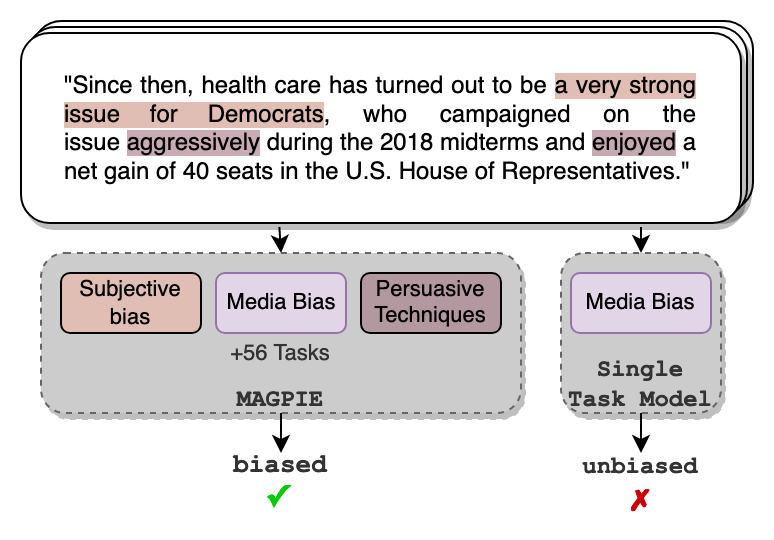}
    \caption{
    \magpie has a pre-trained representation of multiple biases (persuasive, subjective, etc.).
    This enables it to outperform models based on single-task learning (STL) paradigms.
    }
    \label{fig:mtl_demo}
\end{figure}

In this study, we make five main contributions:
\\
\begin{enumerate}

    \item We present \textbf{\magpie} - the first large-scale multi-task pre-training approach for media bias detection.
    By pre-training on diverse bias types such as persuasive and subjective, a classifier based on \magpie correctly classifies sentences that \sotafull single-task models misidentify (we show an example in \Cref{fig:mtl_demo}).\\
    \item We introduce \textbf{\lbm} (\lbmfull), a pre-training composition of 59 bias-related tasks encompassing wide range of biases such as linguistic bias, gender bias and group bias.

    \item We provide an analysis of a task selection and demonstrate the effectiveness of scaling the number of tasks.
    
    \item We demonstrate that \magpie outperforms the previous state-of-the-art model by 3.3\% on the Media Bias Annotation by Experts (BABE) dataset \cite{spinde_neural_2021} and achieves competetive results on the Media Bias Identification Benchmark (MBIB) collection \cite{WesselMBIB22}.

    \item We make all resources, including datasets, training framework, documentation, and models, publicly available on GitHub:
\begin{center} \textbf{\href{https://github.com/Media-Bias-Group/magpie-multi-task}{github.com/magpie-multi-task}\label{magpieurl}}
\end{center}
\end{enumerate} 

These contributions highlight the potential of \mtl in improving media bias detection. 
Our findings show, e.g., that tasks like sentiment and emotionality enhance overall learning, all tasks boost fake news detection, and scaling tasks leads to optimal results.
Another key insight of our research is the value of \mtl in contexts where the primary dataset is small\footnote{For example, the Media Bias Annotation by Experts (BABE) dataset \cite{spinde_neural_2021}.}. 
By learning generalized bias knowledge from a range of tasks, we can improve the accuracy and efficiency of existing models, even in the face of limited data.
Overall, our research offers a multi-task learning approach to media bias detection with first large-scale resources in the domain.

\section{Related Work}\label{sec:RW}
\subsection{Media Bias}
Media bias is a complex issue \cite{lee_mitigating_2021, recasens_linguistic_2013, raza_dbias_2022} composed of varying definitions of bias subtypes such as linguistic bias, context bias, or group bias \cite{WesselMBIB22}. 
In their literature review, \citet{spinde2023media} provide an extensive overview of research on media bias and related subtypes of bias.

Media bias detection approaches have evolved from hand-crafted features \cite{recasens_linguistic_2013, hube_detecting_2018, spinde_automated_2021} to neural models \cite{SpindeJCDL2022, chen_multi-task_2021, spinde_neural_2021, huguet_cabot_us_2021, sinha_determining_2021, raza_dbias_2022}. 
However, existing models, so far, focus only on single tasks and saturate quickly on smaller datasets \cite{WesselMBIB22}. 
As most neural approaches require large quantities of data, those relying on single and small datasets cannot provide a realistic scenario for their solutions (e.g., \citet{fan_plain_2019}). 
We will first provide an overview of existing datasets and then show how to exploit their diversity within the media bias domain.

Media bias tasks and datasets mainly cover individual, self-contained tasks such as binary classifications \cite{recasens_linguistic_2013, spinde_towards_2021}, which, so far, are not explored in relation to each other \cite{spinde2023media}.
\citet{WesselMBIB22} systematically form the media bias detection benchmark MBIB by reviewing over 100 media bias datasets and consolidating 22 of them into eight distinct tasks like linguistic, racial, and political bias. 
Their study highlights that methods only focused on one of these tasks exhibit limitations in their detection capabilities. 
\magpie encompasses all the tasks identified in the MBIB but also significantly expands its scope by incorporating an additional 51 media bias-related tasks to mitigate a variety of limitations in MBIB (see \Cref{sec:lbm}).

\subsection{Multi-Task Learning}
\mtl shows significant improvements in various NLP tasks, including sentiment analysis \cite{he2019interactive}, text summarization \cite{kirstein-etal-2022-analyzing}, and natural language understanding \cite{raffel_exploring_2020}.
In \mtl, a model leverages knowledge gained from one task to improve the performance of others.
\citet{aribandi_ext5_2021} demonstrate that increasing the number of tasks generally leads to improved performance for downstream NLP applications. \citet{aghajanyan-etal-2021-muppet} show that pre-finetuning, a large-scale multi-task learning phase, consistently improves the performance and efficiency of pre-trained models across diverse tasks, with results improving linearly with the number of tasks beyond a certain threshold.

As described above, media bias can be seen as a composite problem composed of various interrelated bias types \cite{WesselMBIB22}. In the realm of Natural Language Understanding (NLU), \mtl has proven to be highly effective when incorporating related tasks \citet{aribandi_ext5_2021}. For instance, benchmarks such as GLUE and SuperGLUE successfully decompose the NLU problem into a suite of proxy tasks, including paraphrase detection \cite{WahleGR23,WahleRKG22a}, and semantic evaluation \cite{RuasGA19,RuasFGF20}, thereby substantially improving performance across a range of NLU tasks \cite{wang-etal-2018-glue, wang_superglue_2019}. Motivated by this success in NLU, we propose to jointly learn from different bias types within the media bias domain. With this approach, we aim to treat media bias not as a singular entity but as many interconnected issues.

The selection of tasks is pivotal to the efficacy of \mtl. 
There have been several attempts to automate task selection, including learning the data selection with Bayesian optimization \cite{bingel-sogaard-2017-identifying} or estimating task relations \cite{ruder-plank-2017-learning}.
The most model-agnostic approach is GradTS \cite{ma-etal-2021-gradts}, which is highly scalable due to low resource requirements, and is therefore implemented within MAGPIE.
GradTS accumulates gradients of attention heads and selects tasks based on their correlation with the primary task's attention.
The selected tasks are trained jointly and share representations across tasks.

\section{Methodolodgy}
We implement \magpie using pre-finetuning as introduced in \cite{aghajanyan-etal-2021-muppet} (See also \Cref{sec:RW}). 
As such, \magpie is an encoder-only \mtl 
transformer model pre-finetuned on 59 media bias-related auxiliary tasks provided by \lbmfull (\lbm), a large-scale task collection of bias-related datasets.
We incorporate a novel approach of a Head-Specific Early Stopping and Resurrection to effectively handle tasks of varying sizes (\Cref{sec:data_sampling}).

\begin{figure}[t]
    \centering
    \includegraphics[width = 0.45\textwidth]{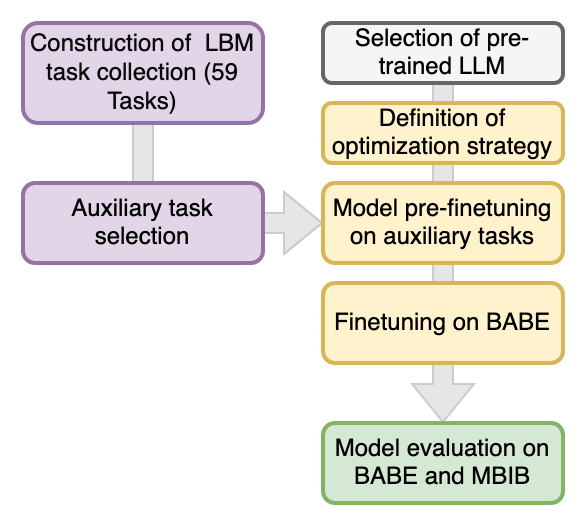}
    \caption{The process of training and evaluating MAGPIE. The purple steps describe the construction and usage of LBM, the yellow the model training, and the green the model evaluation.
    }
    \label{fig:magpie_process}
\end{figure}

As outlined in \Cref{fig:magpie_process}, our first step involves constructing the \lbm. Following this, we define the model and multi-task learning (\mtl) framework employed to train \magpie, which includes optimization strategies, task sampling, and auxiliary task selection. Lastly, we evaluate \magpie on two primary resources: the Media Bias Annotation by Experts (BABE) dataset\footnote{BABE provides high-quality labels that capture a broad range of linguistic biases, thus allowing us to evaluate our model's generalizability within a single dataset context.} \cite{spinde_neural_2021}, and the Media Bias Identification Benchmark (MBIB) collection.

\subsection{The LBM (Large Bias Mixture) task collection} \label{sec:lbm}

\begin{figure}[h]
    \centering
    \includegraphics[width = 0.4\textwidth]{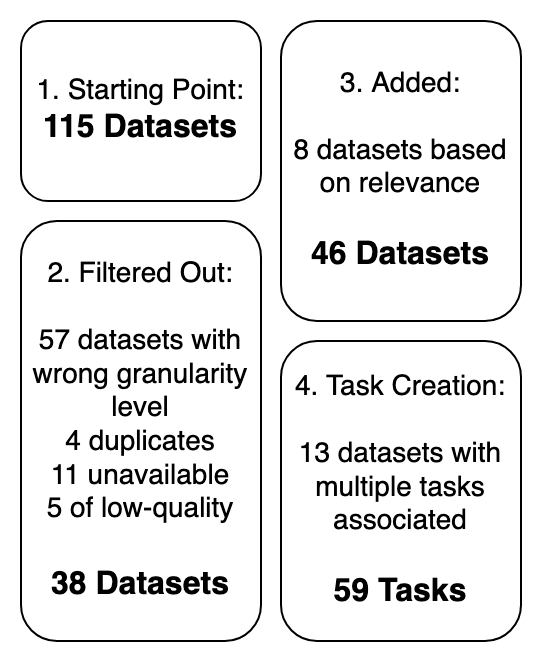}
    \caption{The workflow of collecting datasets from the initial list to the final \lbm collection.}
    \label{fig:lbm_creation}
\end{figure}

Currently, MBIB is the only collection of media bias tasks.
However, it does not include tasks that constitute a form of media bias more indirectly, such as sentiment analysis or emotion classification. 
\magpie aims to integrate tasks both directly linked to media bias and those peripherally related, such as sentiment analysis, to provide broader coverage of linguistic features in the media bias context.
Therefore, we introduce \lbmfull (\lbm), a more general collection of relevant media bias and media bias-related tasks, more suitable for our \mtl approach.
We show our task selection process in \Cref{fig:lbm_creation}.

First, we manually assess a list of 115 media bias-related datasets in English language, categorized into task families by \citet{WesselMBIB22}. 
A task family is a conceptual grouping of tasks that share similar objectives, such as those related to gender bias, encompassing pronoun coreference resolution, gender classification, and sexism detection.

We use this notion of task families to analyze general knowledge transfer between media bias tasks in \Cref{sec:lbmanalysis}, such as \citet{aribandi_ext5_2021} proposed for general NLP tasks.

We filter the collection of the 115 datasets based on the following criteria\footnote{We acknowledge that determining the dataset quality remains a manual and subjective choice.}:
\begin{itemize}
    \item Accessibility: Datasets have to be publicly accessible.
    \item Text granularity: We only use datasets labeled  on a sentence level or its fragments (tokens)(not on, for instance, article level)
    \item Quality of annotations: We exclude datasets with no documentation, low annotation agreement or employment of machine annotation.
    \item Duplicates: We filter out datasets that contain full or partial duplicates of each other
\end{itemize}

Of the 115 datasets collected, we discard 11 datasets that are not publicly available.
We discard 52 with article-level annotations and 5 with annotations on other levels\footnote{One discarded dataset provides only a list of biased words, other two annotations on users, and three on outlets.}.
We remove 5 datasets due to or unreliable source of annotations and discard 4 duplicates. Applying these criteria leaves 38 datasets. Including 8 handpicked datasets not originally listed gives us 46 datasets. These are categorized into task families ensuring no overlap and more than two datasets per family. Finally, ten datasets with multi-level annotations, e.g., token and sentence level, are split into tasks, yielding final number of 59 tasks.

In order to standardize examples from various domains, we apply a unified text-cleaning procedure to each dataset. The process involves: (1) discarding sentences with fewer than 20 characters, (2) removing URL links, special characters, and whitespace, and (3) eliminating duplicates potentially created with steps (1) and (2).

The final \lbmfull (\lbm) includes 59 tasks, categorized into 9 distinct task families, encompassing 1,210,084 labeled sentences. We make the \lbm publicly accessible, to facilitate research in media bias detection and other computational-social-science tasks.
References and short descriptions of all datasets and corresponding tasks and task families can be found in \Cref{table:lbm_all}.

\subsection{The Base Model}

In terms of the base language model for our procedure, we adopt a pre-trained RoBERTa \cite{liu_roberta_2019} encoder due to its proven state-of-the-art performances across various media bias applications \cite{spinde_neural_2021,krieger_domain-adaptive_2022}.

\subsection{The \mtl framework}

\hspace{\parindent}\textbf{Pre-finetuning.}
To effectively harness the generalization abilities of \mtl for media bias detection, we adopt a pre-finetuning procedure \cite{aghajanyan-etal-2021-muppet}.
Pre-finetuning is a complementary approach to pre-training, where a model, already pre-trained on a typically unsupervised task, is subjected to another intermediate step of pre-training.
While incorporating \mtl directly into the pre-training stage has demonstrated performance gains \cite{aribandi_ext5_2021}, we opt for pre-finetuning as it offers significantly reduced computational demands while still capitalizing on the benefits of \mtl \cite{aghajanyan-etal-2021-muppet}.

\textbf{Sharing representations.}
We use hard parameter sharing to share the underlying encoder among all tasks while using separate task-specific heads.
For each task, we attach dense layers, or "heads", to the shared encoder.
These heads are optimized individually per task while the shared encoder learns general bias representations. 

However, multi-task optimization presents challenges due to differing gradient directions and magnitudes \cite{yu2020surgery}. 
For instance, two tasks, A and B, may have opposing gradients with the same magnitude, nullifying their sum.
On the other hand, if Task A's gradient greatly surpasses that of Task B, gradient A becomes dominant. 
We counter the gradient misalignment by using a variation of the PCGrad de-confliction algorithm and loss scaling \cite{yu2020surgery}.

\textbf{Conflicting gradients and loss scaling.}
In multi-task training involving $n$ tasks, encoder parameters receive $n$ potentially conflicting gradients. Efficient handling of this conflict, such as PCGrad \cite{yu2020surgery}, requires storing a set of gradients for each task involved in the update, leading to infeasible memory requirements among our 59 \lbm tasks.
Therefore, we propose a variation of PCGrad we call \textit{PCGrad-online} which preserves the fundamental idea of the original algorithm but is more memory efficient, requiring only one set of gradients instead of $n$ sets per update.
Adopting Muppet's method, we solve the issue of varying task gradient magnitudes by re-scaling the task loss with the inverse log size of its output space, ensuring balanced gradients and preventing task dominance in training steps \cite{aghajanyan-etal-2021-muppet}.

\subsubsection{Data sampling and early stopping}\label{sec:data_sampling}
To prevent large tasks from dominating the optimization, we ensure uniform data distribution by sampling one fixed-size sub-batch from each task per training step, a regular approach in \mtl \cite{aribandi_ext5_2021, SpindeJCDL2022}.
We also employ early stopping as a regularization for each task individually to prevent over-fitting of tasks that converge faster.
However, these methods often fall short when confronted with tasks of varied complexity and differing convergence speeds, which both is the case for tasks in \lbm. 
When task A stops early while task B takes longer to converge, the latent representation of the shared encoder shifts toward task B.

We aim to mitigate this issue by employing a training strategy that tackles the latent representation shift using two complementary approaches:

\begin{enumerate}
    \item Head-Specific-Early-Stopping (HSES)
    \item Resurrection    
\end{enumerate}

\textbf{HSES.} When the task stops, we stop updating its specific head parameters while still backpropagating its language model gradients. This method stems from the observation that not all tasks benefit from the shared layers' continuous learning, especially after they have reached an optimal state.

\textbf{Resurrection.} When the task stops, we allow it to resume the training after its validation loss starts to increase again.
This enables the task to adapt its head parameters to the latest latent representation.

HSES maintains quality of faster-converging tasks, while Resurrection allows further adaptation when needed. Their combination aims for balanced, adaptive learning for tasks with varied complexities and convergence speeds. We perform a preliminary evaluation of the effectiveness of this method in the in \Cref{sec:ablation}. However, we stress the need for further extensive analysis in \Cref{sec:limitations}.

\subsubsection{Auxiliary task selection}\label{subsec:aux}
\mtlfull often involves selecting well-used datasets, leading to potential selection biases. 
Furthermore, manually handpicking datasets becomes challenging due to varying and potentially ambiguous bias annotation schemes.

To automate the process of task selection, we utilize the GradTS algorithm \cite{ma-etal-2021-gradts}.
We choose GradTS due to its demonstrated efficiency and its simplicity of implementation, which enhances its usability.

\textbf{Gradient-based Task selection.}
In line with GradTS, we construct auxiliary tasks as follows. We individually train all tasks, accumulate absolute gradients, then extract and layer-wise normalize these in the attention heads, forming a 12x12 importance matrix\footnote{RoBERTa has 12 attention heads on each of the 12 layers.} for each task. Tasks are sorted by correlation between each task's matrix and BABE task's matrix.

We pre-finetune $m-1$ models on the first $k$ tasks from the sorted list, where $k$ varies from 1 to $m-1$ and $m$ is the size of \lbm. The BABE task is then evaluated on these pre-finetuned models, with the optimal $k$ determined by evaluation loss.
For further details on the GradTS algorithm, please see \citet{ma-etal-2021-gradts}.

\subsubsection{Experimental setup}\label{sec:unstable}
We split the BABE dataset into train, dev and test split with 70\%,15\%,15\% portions respectively. For both hyperparameter tuning and auxiliary task selection, we evaluate on dev split and only use test split for the final evaluation (\Cref{sec:eval}).

As fine-tuning transformers on small datasets often leads to inconsistent results, such as high performance variance \cite{dodge_fine-tuning_2020}, we use a fixed random seed $321$ for all runs in auxilliary task selection.

For the final evaluation on the test set (\Cref{table:results}) we evaluate the models using 30 random seeds and report the average performance, to minimize the influence of random weight initializations. We use values $0..29$.

For optimizing the models, we use a per-task batch size
of 32, an AdamW optimizer, and a polynomial learning scheduler.
We run all experiments on 1 NVIDIA TESLA A100 GPU with 40 GB of memory.

\section{Empirical Results}\label{sec:empirical}
In this section, we present the results of our \mtl approach.
First, we report the set of auxiliary tasks selected by GradTS for pre-finetuning.
Next, we assess how the model pre-finetuned on the GradTS set performs during subsequent finetuning on the BABE task, compared to a random choice of tasks and a full set of \lbm tasks.
We also compare the \mtl approach to a single-task baseline and multiple \mtl baselines and evaluate the performance of our best model, \magpie, on the MBIB benchmark.
Then, we analyze the \lbm taxonomy through a study on knowledge transfer between families.
Lastly, we evaluate the effects of the proposed methods, HSES and Resurrection, through a preliminary study.

\subsection{Auxiliary tasks selection}
We select suitable auxiliary tasks by calculating the correlation between attention-head importance matrices. 
We use Kendall's correlation coefficient, as suggested by \citet{puth_effective_2015}.
We find a local minimum for the BABE evaluation loss when pre-finetuning on the first $k=10$ most correlated tasks.
The final set of the ten most correlated tasks referred to as \textit{gradts} set, is displayed in \Cref{table:gradtsset}.
The tasks in the \textit{gradts} set demonstrate a strong semantic connection to media bias, encompassing areas such as lexical bias, rumor detection, and fake news detection.

\begin{table}
\begin{tabular}{|c|c|}
\hline
Task Type & $\tau$ (correlation) \\
\hline
\hline
Persuasive techniques & 0.73 \\
Lexical/Informational bias & 0.72 \\
Rumour detection & 0.69 \\
Sentiment analysis & 0.68 \\
Global warming stance detection & 0.68 \\
Subjective bias & 0.67 \\
Veracity classification & 0.64 \\
Gender bias & 0.64 \\
Fake news detection & 0.63 \\
\hline
\end{tabular}
\caption{Attention importance correlation w.r.t the media bias task.}
\label{table:gradtsset}
\end{table}

\newpage
\subsection{Evaluation}\label{sec:eval}
First, we finetune a model pre-finetuned on three different multi-task sets on the BABE task and compare it against the single-task RoBERTa baseline.
The multi-task sets are the following:

\begin{itemize}
    \item \textbf{MTL:Random} - Random Subset of 10 tasks
    \item  \textbf{MTL:GradTS} - Subset of 10 tasks selected by GradTS algorithm
    \item \textbf{MTL:All} - Set of all tasks
\end{itemize}

We also evaluate the model pre-finetuned on the set of all tasks on MBIB. 
We follow the guidelines set by \citet{WesselMBIB22} for the evaluation.
Given that \magpie's pre-training data includes portions of the MBIB data, we ensure that the test set for each task in MBIB is not exposed to the model during its training or validation phases.

\begin{figure}[h]
    \centering
    \includegraphics[width = 0.5\textwidth]{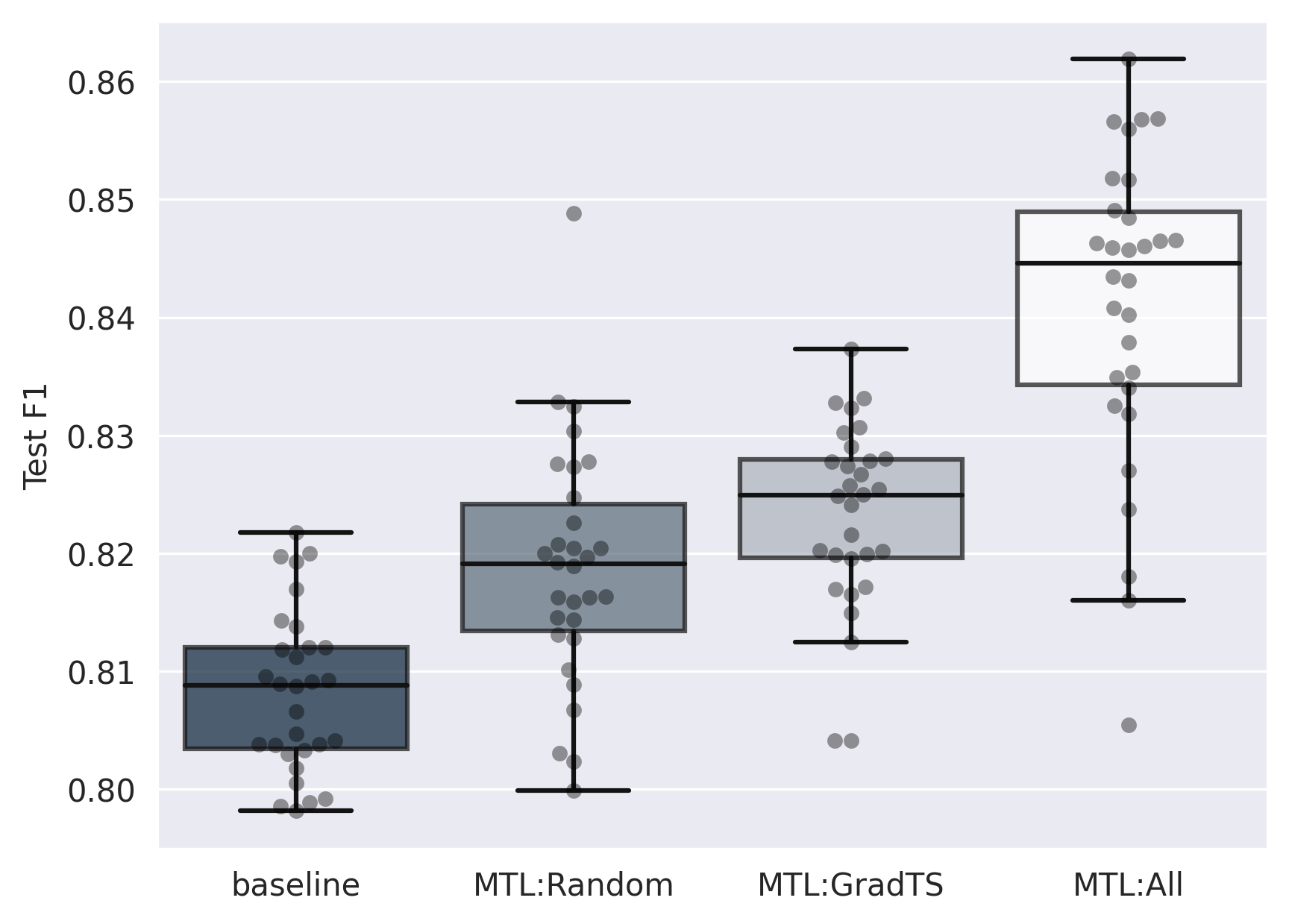}
    \caption{Final F1 score on a BABE test set averaged over 30 random seeds. All three \mtl approaches outperform baseline STL finetuning. Pre-finetuning on all \lbm tasks results in significantly improved performance.}
    \label{fig:test_f1}
\end{figure}

\textbf{Multi-Task Learning Performance.} \Cref{table:results} summarizes our performance results on the BABE dataset.
We observe that all of our \mtl pre-finetuning schemas lead to performance improvements.
In particular, pre-finetuning on \textbf{all} tasks from \lbm yields a \sota performance on the BABE dataset, achieving an 84.1\% F1 score and a relative improvement of 3.3\% compared to the previous baseline by \citet{spinde_neural_2021}.
While both \mtl baselines - Muppet \cite{aghajanyan-etal-2021-muppet} and UnifiedM2 \cite{lee-etal-2021-unifying} outperform single-task baseline, they underperform all of our \mtl models.

On MBIB benchmark, MAGPIE ranks first on 5 out of 8 tasks.
However, the improvements are only marginal. The results can be found in Appendix in \Cref{table:MBIB_comparison}.
\\

\textbf{Task scaling.}
GradTS task selection outperforms random tasks on average performance, yet our experiment suggests task number scaling is more crucial. This is consistent with Muppet and ExT5 results \cite{aghajanyan-etal-2021-muppet,aribandi_ext5_2021}, indicating \mtl can compensate for scarce high-quality media bias datasets through general bias representation from other tasks. It also supports \citet{kirstein-etal-2022-analyzing}'s finding that sufficient related tasks can substitute the original task.

\begin{center}
\begin{table*}[ht]
\centering
\begin{tabular}{|l|l|l|l|}
\hline
                    \textbf{Model} &            \textbf{F1 } &           \textbf{Acc} &          \textbf{loss} \\
\hline
\hline
                 Baseline (RoBERTa base) & 80.83 (±0.69) & 81.19 (±0.69) &  43.6 (±3.54) \\
               DA-RoBERTa &  77.83 (±1.4) &  78.56 (±1.3) & 47.84 (±2.97) \\
                   MUPPET &  80.56 (±1.3) & 81.18 (±1.16) & 44.19 (±4.65) \\
UnifiedM2 & 81.91 (±0.91) & 82.41 (±0.88) & 44.86 (±3.99) \\
               MTL:Random & 81.88 (±1.02) & 82.28 (±0.97) & 40.35 (±1.73) \\
               MTL:GradTS & 82.32 (±0.79) &  82.64 (±0.8) & 40.96 (±2.36) \\
                  MTL:All &  \textbf{84.1} (±1.33) & \textbf{84.44} (±1.25) & \textbf{39.46} (±2.41) \\
\hline
\end{tabular}
\caption{Performance of two \mtl baseline models (Muppet, UnifiedM2) two single-task baslines (RoBERTa and DA-RoBERTa) and our three \mtl models, on fine-tuning on BABE dataset and evaluating on the held-out test set. The results are averaged over 30 random seeds.}
\label{table:results}

\end{table*}
\end{center}

\begin{figure}[h]
    \centering
    \includegraphics[width = 0.5\textwidth]{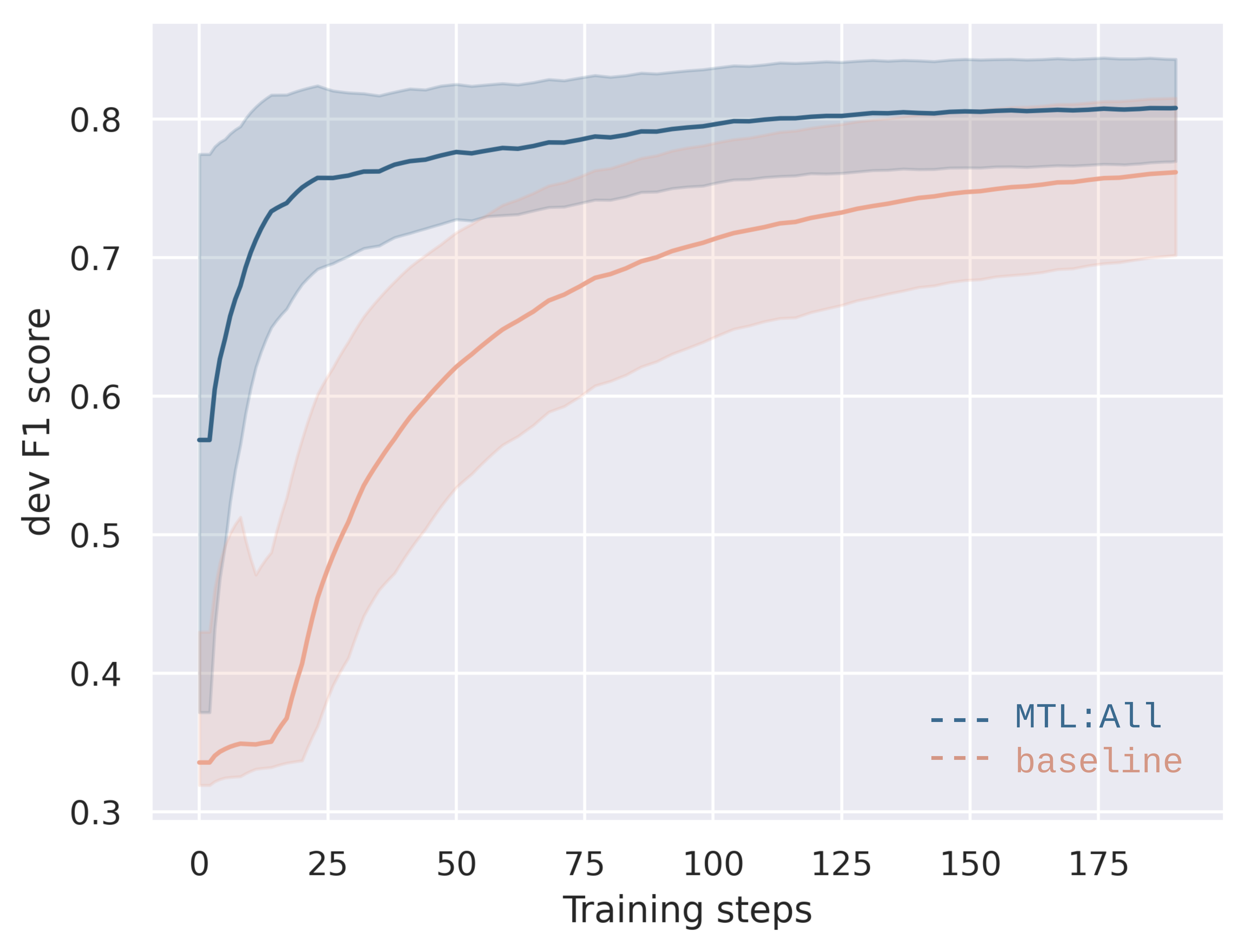}
    \caption{Evaluation F1 score during the final finetuning where \mtl: All shows superior performance in training-step efficiency. The values are averaged over 30 random seeds. The bands mark the lowest and highest values.}
    \label{fig:f1-curve}
\end{figure}

\textbf{Step efficiency.}
In addition to the performance improvements achieved, we also assess our model training efficiency. 
In \Cref{fig:f1-curve}, we show the F1 score on the development set for the BABE task, averaged over all 30 runs.
Our findings show that \mtlfull only requires \~15\% of the training steps used in single-task finetuning on BABE. 
This result demonstrates the high training-step efficiency of \magpie in media bias classification, making \mtl implementations in the media bias domain more viable in the future.

\subsection{LBM taxonomy analysis}
\label{sec:lbmanalysis}
In \Cref{sec:lbm}, following \citet{aribandi_ext5_2021}, we introduce data task families. \citet{aribandi_ext5_2021} uses task families for selection and knowledge transfer.
To assess task families' significance in \lbm taxonomy, we train each pair of families together, investigating knowledge transfer.

\begin{figure}[h]
    \centering
    \includegraphics[width=0.5\textwidth]{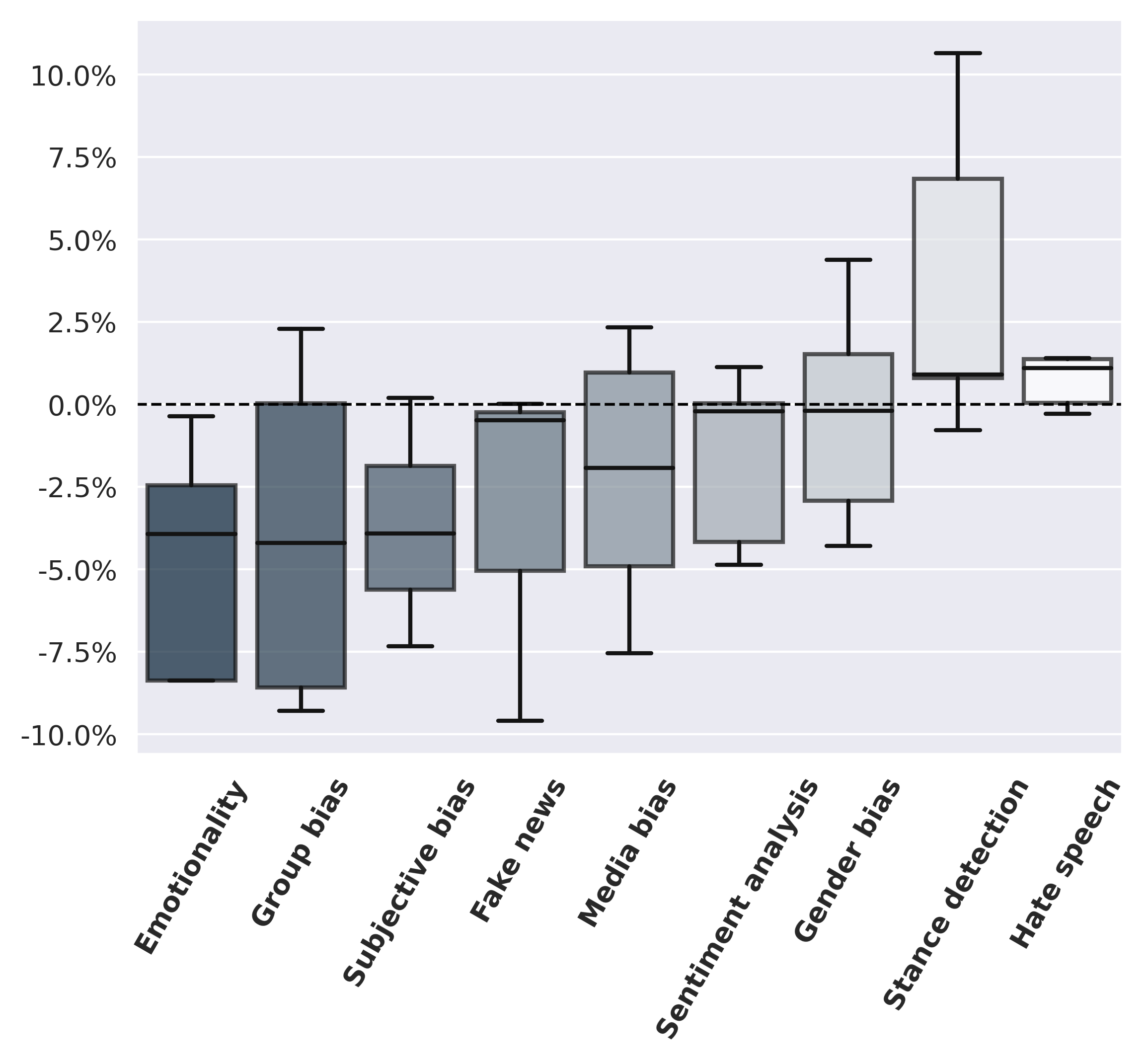}
    \caption{Average performance change per task family. Stance detection and hate speech are the only families, on average, benefitting from \mtlfull.}
    \label{fig:withinfamily}
\end{figure}

To account for potential negative transfer within families, we first calculate the average transfer within each family and use it as a baseline for measuring transfer between families.
We train tasks from the same family together and report the average change in task performance,  as depicted in \Cref{fig:withinfamily}.
Negative knowledge transfer is prevalent across most of our task families. 
However, we observe two exceptions: the hate-speech and stance detection families, where multi-task training leads to an average improvement in performance.

Next, we measure knowledge transfer between families by training each pair together. We report the average impact of each family on others and the average benefit each family gets through training with others, summarized in \Cref{table:between_family_transfer_evaluation}.

Our results show that, on average, the Emotionality and Sentiment analysis families provide positive transfer learning to other families.
Conversely, we observe that the Fake News family benefits from knowledge transfer from every other family, with an average improvement of 1.7\%.
On the other hand, the Emotionality family is significantly impaired by negative transfer from other task families.

\begin{figure*}
    \centering
    \includegraphics[width =\textwidth]{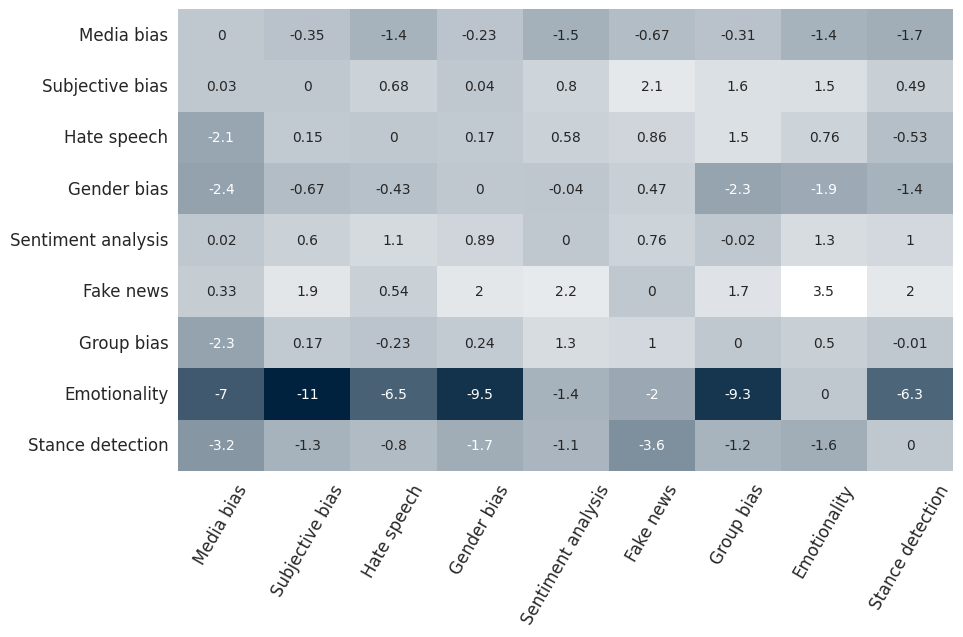}

    \caption{Average performance change when training tasks together. The change is measured with respect to transfer within each respective family (see results in section \Cref{fig:withinfamily}.). The values in the horizontal axis represent the received transfer by the family on the y-axis. E.g., when training Emotionality and Subjective bias together, Emotionality gets worse by 11\% whereas Subjectivity improves by 1.5\%.}
    \label{fig:bft}
\end{figure*}

The full table of transfers can be found in \Cref{fig:bft}.
Considering that only two families show positive transfer learning, with marginal effects of 0.11\% and 0.34\%, we conclude that the task families used in the construction of \lbm are generally unsuitable for effectively utilizing knowledge transfer. We discuss this again in \Cref{sec:final}.

\begin{table}
\label{table:between_family_transfer_evaluation}
\begin{tabular}{|c|c|c|}
\hline
Task Family & Transfer \textbf{from} & Transfer \textbf{to} \\
\hline
\hline
media bias & \cellcolor{red}{-2.07}\% & -0.94\% \\
subjective bias & -1.26\% & 0.89\% \\
hate speech & -0.87\% & 0.17\% \\
gender bias & -1.01\% & -1.07\% \\
sentiment analysis & \cellcolor{green}{0.11}\% & 0.72\% \\
fake news & -0.13\% & \cellcolor{green}{1.79}\% \\
group bias & -1.04\% & 0.09\% \\
emotionality & \cellcolor{green}{0.34}\% & \cellcolor{red}{-6.56}\% \\
stance detection & -0.79\% & -1.83\% \\
\hline
\end{tabular}
\caption{Evaluation of averaged transfer between task families.}

\end{table}

\subsection{Resurrection and HSES evaluation}\label{sec:ablation}

\begin{figure}
    \subfloat[\centering Resurrection]{{\includegraphics[width=0.23\textwidth]{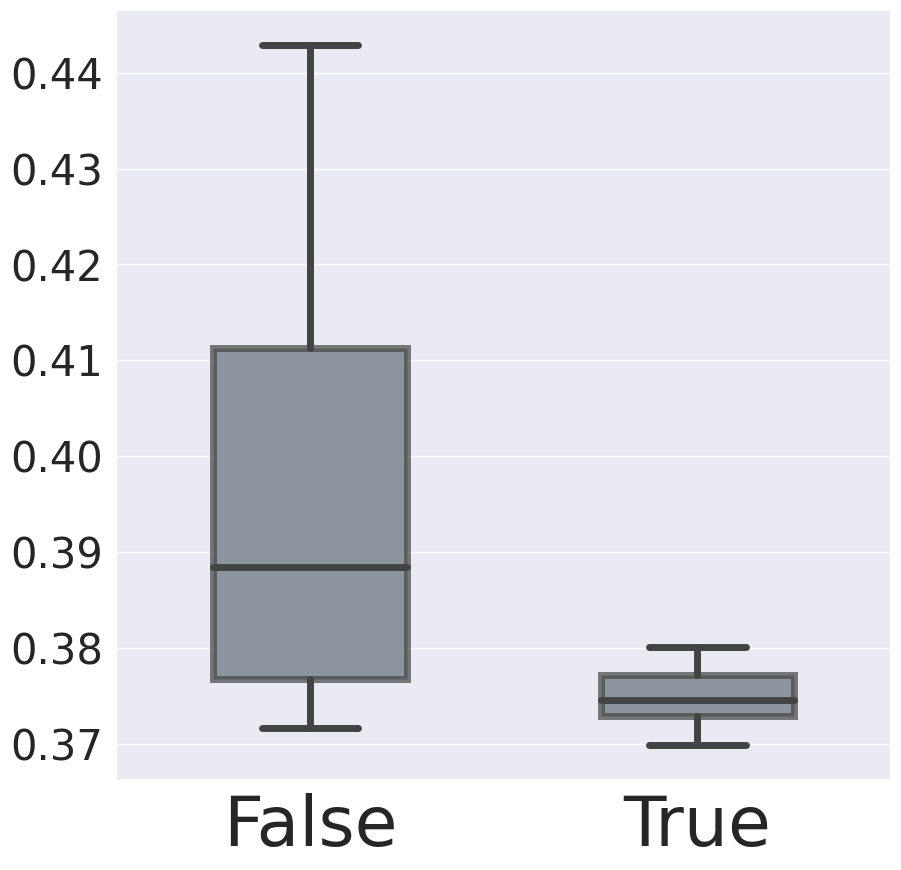} }}\hfill
    \subfloat[\centering HSES]{{\includegraphics[width=0.23\textwidth]{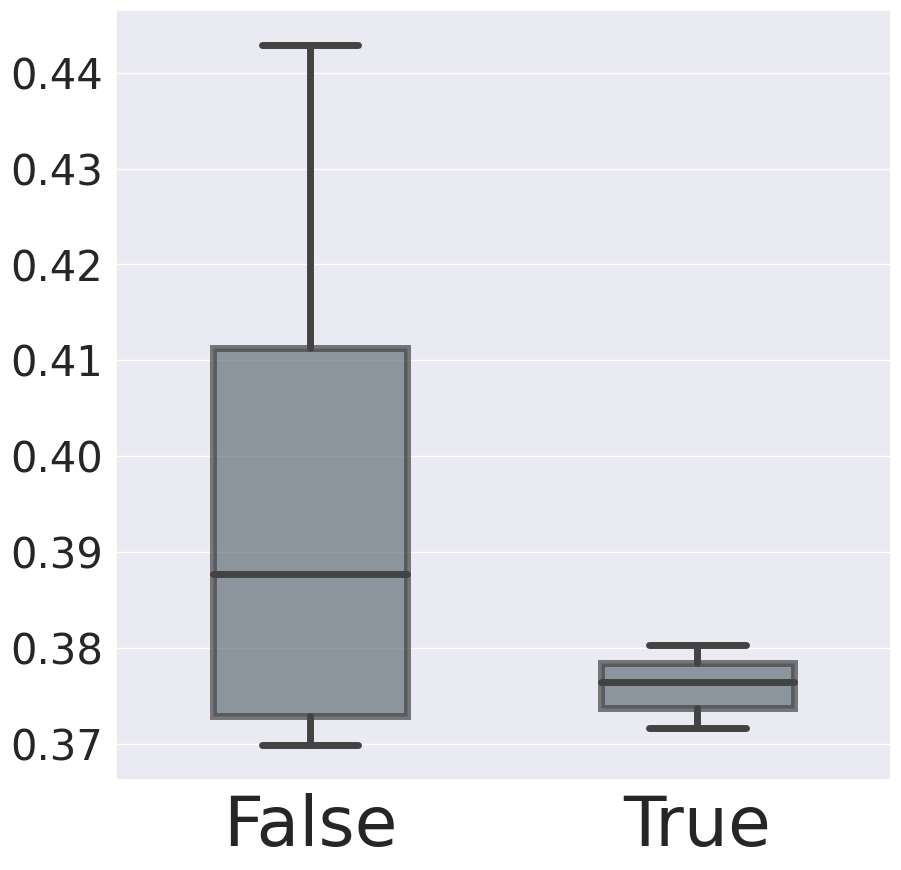} }}
    \caption{An averaged evaluation loss of 20 tasks when trained with different training strategies. Both Resurrection and HSES approaches, compared to the vanilla setting, lead to significantly lower variance and overall lower loss.}
    \label{fig:ablation}
\end{figure}

To evaluate the Resurrection and HSES methods in combination with other training strategies, we run a grid search on the following training strategies: \textit{HSES}, \textit{Resurrection}, \textit{Loss Scaling} and \textit{Gradient Aggregation}.
We calculate the average evaluation loss for both Resurrection and HSES methods across 20 tasks randomly selected from the \lbm collection. 
The boxplot in \Cref{fig:ablation} shows that both methods reduce the loss by 5\% and decrease the variance across different training setups by 85\%.
However, we hypothesize that a random constellation of tasks\footnote{Particularly variance in task sizes and quality.} can have a non-trivial effect on the evaluation of our technique; thus, we opt for robust examination of the methods in future work.

\section{Conclusion} \label{sec:final}

This paper contributes to media bias detection by the development of \magpie, our large-scale multi-task learning (\mtl) approach that enhances traditional models with a multi-task structure.

Additionally, we present \lbmfull (\lbm), a  compilation of 59 bias-related tasks. This broad collection serves as a resource for the pre-training of \magpie. To the best of our knowledge, it is the first available \mtl resource tailored specifically towards media bias. 

Our study investigates the dynamics of transfer learning across tasks and task families within media bias. Despite the occurrence of negative transfer among several tasks, scaling the pre-training setup to all collected tasks in \mtlfull results in a 3.3\% improvement over the previous \sotafull, making it the biggest advancement in neural media bias classification so far.
Furthermore, we report that finetuning \magpie on the BABE dataset only requires 15\% of steps compared to RoBERTa single-task approaches.
These findings underscore the effectiveness and potency of \mtlfull in highly specific classification domains such as media bias.

While results suggest benefits in scaling tasks, we see more promise in novel tasks rooted in media bias, suggesting deeper exploration over simply expanding the task spectrum.
Understanding families and tasks in datasets necessitates systematic analysis of label definitions, rater agreement, and inter-relatedness of dataset creation strategies.
As media bias is emerging globally, incorporating multilingual models is a natural extension.

\section*{Limitations}\label{sec:limitations}
We acknowledge the necessity for a more comprehensive analysis of the performance of the HSES and Resurrection methods to ensure robust evaluation.

Given the significant computational resources required for a single multi-task training of all tasks, we allocated these resources towards robustly evaluating the model performance rather than conducting an in-depth analysis of the optimization techniques. Consequently, the presented methods may be less reliable and have limited applicability.

While there are various models available, previous research suggests that RoBERTa exhibits strong performance on media bias tasks \cite{WesselMBIB22, spinde_neural_2021}. However, due to resource constraints, we were unable to explore models with different architectures and further refine our selection.

As the landscape of publicly available high-quality datasets for media bias is not as extensive as desired, we acknowledge our inability to capture all manifestations of media bias comprehensively. As mentioned in \Cref{sec:final}, a systematic and comprehensive analysis of the dataset landscape will be part of our future work. 

Furthermore, although analyzing media bias on a sentence level enables a detailed examination of occurring biases, certain forms of bias, such as bias resulting from the omission of information, extend beyond the linguistic aspects of individual statements. Detecting such biases may require considering an external context.

We leave it to future work to investigate other levels of granularity, such as paragraph or article levels. In addition to these technical limitations, conducting a detailed analysis of agreement and label distribution in all utilized datasets will be necessary to make stronger claims about which datasets provide more reliable coverage of the overall concept of media bias. This is particularly important since media bias is a complex phenomenon that is not always easily identified during the annotation-gathering process.

\section*{Ethics Statement}
Detecting (and as a result also highlighting) media bias instances can mitigate the negative effects of media bias, e.g., on collective decision-making \cite{baumer_testing_2015}. However, media bias strongly depends on the context and individuals involved, making it a sensitive issue. Some bias forms depend on factors other than the content, e.g., a different text perception due to a reader’s background. Therefore, a public classification of possible polarization and one-sidedness in the news must be performed transparently and reliably. The datasets used for training the classifier must be transparent and, ideally, constantly monitored by independent experts. To address possible implications of \magpie or related media bias applications cautiously, it is crucial to constantly control the classifications and datasets with recent findings on the perception of media bias, which is a main part of our ongoing and future work. To do so, we use resources such as, e.g., standardized questionnaires on media bias perception \cite{spinde_you_2021}.

When bias detection is balanced and transparent, it can positively affect collective opinion formation and decision-making. We see no immediate negative ethical or societal impacts of our work. However, in addition to system transparency, we want to highlight that we believe it to be required for future datasets to report in greater detail about a manipulation protection strategy when developing, training, and presenting any media bias classifier. To ensure the validity of media bias detection systems, it is essential to prevent participants, especially in public studies, from manipulating algorithms by, for example, flagging neutral content as biased. Therefore, annotations should be compared among multiple users to ensure a higher level of trustworthiness. Most of the datasets available report only limitedly about such strategies. In open or crowdsourcing scenarios, it is important to collect user characteristics and deliberately include specific content that aims to provide obvious answers but may be answered differently when users follow certain patterns. This approach helps in detecting and mitigating potential biases introduced by participants, thereby maintaining the integrity and reliability of the media bias detection process. To ensure and propose stronger standards in the future, we aim to analyze all \lbm datasets with regard to potential inherent bias in future work.

\section*{Acknowledgments}
We are grateful for the financial support of this project provided by the Hanns-Seidel Foundation, the DAAD (German Academic Exchange Service, program number 57515303 and 57515245), the Lower Saxony Ministry of Science and Culture, the VW Foundation and the AI Center of Czech Technical University. Furthermore, the authors would like to express their gratitude towards Jan Drchal for administrative support.

\nocite{*}
\section{Bibliographical References}\label{sec:reference}

\bibliographystyle{lrec-coling2024-natbib}
\bibliography{custom_bib}

\begin{thebibliography}{114}
\expandafter\ifx\csname natexlab\endcsname\relax\def\natexlab#1{#1}\fi

\bibitem[{Aghajanyan et~al.(2021)Aghajanyan, Gupta, Shrivastava, Chen, Zettlemoyer, and Gupta}]{aghajanyan-etal-2021-muppet}
Armen Aghajanyan, Anchit Gupta, Akshat Shrivastava, Xilun Chen, Luke Zettlemoyer, and Sonal Gupta. 2021.
\newblock \href {https://doi.org/10.18653/v1/2021.emnlp-main.468} {Muppet: Massive multi-task representations with pre-finetuning}.
\newblock In \emph{Proceedings of the 2021 Conference on Empirical Methods in Natural Language Processing}, pages 5799--5811, Online and Punta Cana, Dominican Republic. Association for Computational Linguistics.

\bibitem[{AI(2019)}]{jigsaw/conversationaiJigsawUnintendedBias2019}
Jigsaw/Conversation AI. 2019.
\newblock \href {https://www.kaggle.com/competitions/jigsaw-unintended-bias-in-toxicity-classification/data} {Jigsaw unintended bias in toxicity classification.}

\bibitem[{ALDayel and Magdy(2021)}]{aldayelStanceDetectionSocial2021}
Abeer ALDayel and Walid Magdy. 2021.
\newblock \href {https://doi.org/10.1016/j.ipm.2021.102597} {Stance detection on social media: {State} of the art and trends}.
\newblock \emph{Information Processing \& Management}, 58(4):102597.

\bibitem[{Aribandi et~al.(2021{\natexlab{a}})Aribandi, Tay, and Metzler}]{aribandi-etal-2021-reliable}
Vamsi Aribandi, Yi~Tay, and Donald Metzler. 2021{\natexlab{a}}.
\newblock \href {https://doi.org/10.18653/v1/2021.findings-acl.155} {How reliable are model diagnostics?}
\newblock In \emph{Findings of the Association for Computational Linguistics: ACL-IJCNLP 2021}, pages 1778--1785, Online. Association for Computational Linguistics.

\bibitem[{Aribandi et~al.(2021{\natexlab{b}})Aribandi, Tay, Schuster, Rao, Zheng, Mehta, Zhuang, Tran, Bahri, Ni, Gupta, Hui, Ruder, and Metzler}]{aribandi_ext5_2021}
Vamsi Aribandi, Yi~Tay, Tal Schuster, Jinfeng Rao, Huaixiu~Steven Zheng, Sanket~Vaibhav Mehta, Honglei Zhuang, Vinh~Q. Tran, Dara Bahri, Jianmo Ni, Jai~Prakash Gupta, Kai Hui, Sebastian Ruder, and Donald Metzler. 2021{\natexlab{b}}.
\newblock \href {https://arxiv.org/abs/2111.10952} {{ExT5}: {Towards} {Extreme} {Multi}-{Task} {Scaling} for {Transfer} {Learning}}.
\newblock \emph{CoRR}, abs/2111.10952.
\newblock ArXiv: 2111.10952.

\bibitem[{Attree(2019)}]{attreeGenderedAmbiguousPronouns2019}
Sandeep Attree. 2019.
\newblock \href {https://doi.org/10.48550/ARXIV.1906.00839} {Gendered {Ambiguous} {Pronouns} {Shared} {Task}: {Boosting} {Model} {Confidence} by {Evidence} {Pooling}}.

\bibitem[{Barikeri et~al.(2021{\natexlab{a}})Barikeri, Lauscher, Vulić, and Glavaš}]{barikeriRedditBiasRealWorldResource2021}
Soumya Barikeri, Anne Lauscher, Ivan Vulić, and Goran Glavaš. 2021{\natexlab{a}}.
\newblock \href {https://doi.org/10.48550/ARXIV.2106.03521} {{RedditBias}: {A} {Real}-{World} {Resource} for {Bias} {Evaluation} and {Debiasing} of {Conversational} {Language} {Models}}.

\bibitem[{Barikeri et~al.(2021{\natexlab{b}})Barikeri, Lauscher, Vulić, and Glavaš}]{barikeri_redditbias_2021}
Soumya Barikeri, Anne Lauscher, Ivan Vulić, and Goran Glavaš. 2021{\natexlab{b}}.
\newblock \href {https://doi.org/10.18653/v1/2021.acl-long.151} {{RedditBias}: {A} {Real}-{World} {Resource} for {Bias} {Evaluation} and {Debiasing} of {Conversational} {Language} {Models}}.
\newblock In \emph{Proceedings of the 59th {Annual} {Meeting} of the {Association} for {Computational} {Linguistics} and the 11th {International} {Joint} {Conference} on {Natural} {Language} {Processing} ({Volume} 1: {Long} {Papers})}, pages 1941--1955, Online. Association for Computational Linguistics.

\bibitem[{Baumer et~al.(2015)Baumer, Elovic, Qin, Polletta, and Gay}]{baumer_testing_2015}
Eric Baumer, Elisha Elovic, Ying Qin, Francesca Polletta, and Geri Gay. 2015.
\newblock \href {https://doi.org/10.3115/v1/N15-1171} {Testing and {Comparing} {Computational} {Approaches} for {Identifying} the {Language} of {Framing} in {Political} {News}}.
\newblock In \emph{Proceedings of the 2015 {Conference} of the {North} {American} {Chapter} of the {Association} for {Computational} {Linguistics}: {Human} {Language} {Technologies}}, pages 1472--1482, Denver, Colorado. Association for Computational Linguistics.

\bibitem[{Bingel and S{\o}gaard(2017)}]{bingel-sogaard-2017-identifying}
Joachim Bingel and Anders S{\o}gaard. 2017.
\newblock \href {https://aclanthology.org/E17-2026} {Identifying beneficial task relations for multi-task learning in deep neural networks}.
\newblock In \emph{Proceedings of the 15th Conference of the {E}uropean Chapter of the Association for Computational Linguistics: Volume 2, Short Papers}, pages 164--169, Valencia, Spain. Association for Computational Linguistics.

\bibitem[{Bostan et~al.(2020)Bostan, Kim, and Klinger}]{bostanGoodNewsEveryoneCorpusNews2020}
Laura Ana~Maria Bostan, Evgeny Kim, and Roman Klinger. 2020.
\newblock \href {https://aclanthology.org/2020.lrec-1.194} {{GoodNewsEveryone}: {A} {Corpus} of {News} {Headlines} {Annotated} with {Emotions}, {Semantic} {Roles}, and {Reader} {Perception}}.
\newblock In \emph{Proceedings of the 12th {Language} {Resources} and {Evaluation} {Conference}}, pages 1554--1566, Marseille, France. European Language Resources Association.

\bibitem[{BSI(1973{\natexlab{a}})}]{bs-2570-manual}
BSI. 1973{\natexlab{a}}.
\newblock \emph{Natural Fibre Twines}, 3rd edition.
\newblock British Standards Institution, London.
\newblock BS 2570.

\bibitem[{BSI(1973{\natexlab{b}})}]{bs-2570-techreport}
BSI. 1973{\natexlab{b}}.
\newblock Natural fibre twines.
\newblock BS 2570, British Standards Institution, London.
\newblock 3rd. edn.

\bibitem[{Castor and Pollux(1992)}]{CastorPollux-92}
A.~Castor and L.~E. Pollux. 1992.
\newblock The use of user modelling to guide inference and learning.
\newblock \emph{Applied Intelligence}, 2(1):37--53.

\bibitem[{Chen et~al.(2021)Chen, Zhang, and Yang}]{chen_multi-task_2021}
Shijie Chen, Yu~Zhang, and Qiang Yang. 2021.
\newblock \href {http://arxiv.org/abs/2109.09138} {Multi-{Task} {Learning} in {Natural} {Language} {Processing}: {An} {Overview}}.
\newblock \emph{arXiv:2109.09138 [cs]}.

\bibitem[{Chercheur(1994)}]{Chercheur-94}
J.L. Chercheur. 1994.
\newblock \emph{Case-Based Reasoning}, 2nd edition.
\newblock Morgan Kaufman Publishers, San Mateo, CA.

\bibitem[{Chomsky(1973)}]{chomsky-73}
N.~Chomsky. 1973.
\newblock Conditions on transformations.
\newblock In \emph{A festschrift for {Morris Halle}}, New York. Holt, Rinehart \& Winston.

\bibitem[{Conforti et~al.(2020)Conforti, Berndt, Pilehvar, Giannitsarou, Toxvaerd, and Collier}]{confortiWillTheyWonTTheyVery2020}
Costanza Conforti, Jakob Berndt, Mohammad~Taher Pilehvar, Chryssi Giannitsarou, Flavio Toxvaerd, and Nigel Collier. 2020.
\newblock \href {https://doi.org/10.18653/v1/2020.acl-main.157} {Will-they-won{'}t-they: A very large dataset for stance detection on {T}witter}.
\newblock In \emph{Proceedings of the 58th Annual Meeting of the Association for Computational Linguistics}, pages 1715--1724, Online. Association for Computational Linguistics.

\bibitem[{Dacon and Liu(2021)}]{daconDoesGenderMatter2021}
Jamell Dacon and Haochen Liu. 2021.
\newblock \href {https://doi.org/10.1145/3442442.3452325} {Does {Gender} {Matter} in the {News}? {Detecting} and {Examining} {Gender} {Bias} in {News} {Articles}}.
\newblock In \emph{Companion {Proceedings} of the {Web} {Conference} 2021}, {WWW} '21, pages 385--392, New York, NY, USA. Association for Computing Machinery.

\bibitem[{Darwish et~al.(2017)Darwish, Magdy, and Zanouda}]{darwishTrumpVsHillary2017}
Kareem Darwish, Walid Magdy, and Tahar Zanouda. 2017.
\newblock \href {https://doi.org/10.1007/978-3-319-67217-5_10} {Trump vs. {Hillary}: {What} {Went} {Viral} {During} the 2016 {US} {Presidential} {Election}}.
\newblock In \emph{Social {Informatics}}, Lecture {Notes} in {Computer} {Science}, pages 143--161, Cham. Springer International Publishing.

\bibitem[{Davidson et~al.(2017)Davidson, Warmsley, Macy, and Weber}]{davidson_automated_2017}
Thomas Davidson, Dana Warmsley, Michael Macy, and Ingmar Weber. 2017.
\newblock \href {https://doi.org/https://doi.org/10.1609/icwsm.v11i1.14955} {Automated hate speech detection and the problem of offensive language}.
\newblock In \emph{Proceedings of the {International} {AAAI} {Conference} on {Web} and {Social} {Media}}, volume~11.
\newblock Issue: 1.

\bibitem[{Dinan et~al.(2020)Dinan, Fan, Wu, Weston, Kiela, and Williams}]{dinanMultiDimensionalGenderBias2020}
Emily Dinan, Angela Fan, Ledell Wu, Jason Weston, Douwe Kiela, and Adina Williams. 2020.
\newblock \href {https://doi.org/10.18653/v1/2020.emnlp-main.23} {Multi-dimensional gender bias classification}.
\newblock In \emph{Proceedings of the 2020 Conference on Empirical Methods in Natural Language Processing (EMNLP)}, pages 314--331, Online. Association for Computational Linguistics.

\bibitem[{Dinan et~al.(2018)Dinan, Roller, Shuster, Fan, Auli, and Weston}]{dinanWizardWikipediaKnowledgePowered2018}
Emily Dinan, Stephen Roller, Kurt Shuster, Angela Fan, Michael Auli, and Jason Weston. 2018.
\newblock \href {https://doi.org/10.48550/ARXIV.1811.01241} {Wizard of {Wikipedia}: {Knowledge}-{Powered} {Conversational} agents}.

\bibitem[{Dinan et~al.(2019)Dinan, Roller, Shuster, Fan, Auli, and Weston}]{dinan_wizard_2019}
Emily Dinan, Stephen Roller, Kurt Shuster, Angela Fan, Michael Auli, and Jason Weston. 2019.
\newblock Wizard of {Wikipedia}: {Knowledge}-powered {Conversational} {Agents}.
\newblock In \emph{Proceedings of the {International} {Conference} on {Learning} {Representations} ({ICLR})}.

\bibitem[{Dixon et~al.(2018)Dixon, Li, Sorensen, Thain, and Vasserman}]{dixonMeasuringMitigatingUnintended2018}
Lucas Dixon, John Li, Jeffrey Sorensen, Nithum Thain, and Lucy Vasserman. 2018.
\newblock \href {https://doi.org/10.1145/3278721.3278729} {Measuring and {Mitigating} {Unintended} {Bias} in {Text} {Classification}}.
\newblock In \emph{Proceedings of the 2018 {AAAI}/{ACM} {Conference} on {AI}, {Ethics}, and {Society}}, {AIES} '18, pages 67--73, New York, NY, USA. Association for Computing Machinery.

\bibitem[{Dodge et~al.(2020)Dodge, Ilharco, Schwartz, Farhadi, Hajishirzi, and Smith}]{dodge_fine-tuning_2020}
Jesse Dodge, Gabriel Ilharco, Roy Schwartz, Ali Farhadi, Hannaneh Hajishirzi, and Noah Smith. 2020.
\newblock \href {https://doi.org/10.48550/ARXIV.2002.06305} {Fine-{Tuning} {Pretrained} {Language} {Models}: {Weight} {Initializations}, {Data} {Orders}, and {Early} {Stopping}}.

\bibitem[{Eco(1990)}]{Eco:1990}
Umberto Eco. 1990.
\newblock \emph{The Limits of Interpretation}.
\newblock Indian University Press.

\bibitem[{Fan et~al.(2019)Fan, White, Sharma, Su, Choubey, Huang, and Wang}]{fan_plain_2019}
Lisa Fan, Marshall White, Eva Sharma, Ruisi Su, Prafulla~Kumar Choubey, Ruihong Huang, and Lu~Wang. 2019.
\newblock \href {https://doi.org/10.18653/v1/D19-1664} {In plain sight: Media bias through the lens of factual reporting}.
\newblock In \emph{Proceedings of the 2019 Conference on Empirical Methods in Natural Language Processing and the 9th International Joint Conference on Natural Language Processing (EMNLP-IJCNLP)}, pages 6343--6349, Hong Kong, China. Association for Computational Linguistics.

\bibitem[{Ferreira and Vlachos(2016)}]{ferreiraEmergentNovelDataset2016}
W.~Ferreira and A.~Vlachos. 2016.
\newblock \href {http://aclweb.org/anthology/N/N16/N16-1138.pdf} {Emergent: a novel data-set for stance classification}.

\bibitem[{Founta et~al.(2018)Founta, Djouvas, Chatzakou, Leontiadis, Blackburn, Stringhini, Vakali, Sirivianos, and Kourtellis}]{fountaLargeScaleCrowdsourcing2018}
Antigoni Founta, Constantinos Djouvas, Despoina Chatzakou, Ilias Leontiadis, Jeremy Blackburn, Gianluca Stringhini, Athena Vakali, Michael Sirivianos, and Nicolas Kourtellis. 2018.
\newblock \href {https://doi.org/10.1609/icwsm.v12i1.14991} {Large {Scale} {Crowdsourcing} and {Characterization} of {Twitter} {Abusive} {Behavior}}.
\newblock \emph{Proceedings of the International AAAI Conference on Web and Social Media}, 12(1).
\newblock Number: 1.

\bibitem[{Färber et~al.(2020)Färber, Burkard, Jatowt, and Lim}]{farber_multidimensional_2020}
Michael Färber, Victoria Burkard, Adam Jatowt, and Sora Lim. 2020.
\newblock \href {https://doi.org/10.1145/3340531.3412876} {A {Multidimensional} {Dataset} {Based} on {Crowdsourcing} for {Analyzing} and {Detecting} {News} {Bias}}.
\newblock In \emph{Proceedings of the 29th {ACM} {International} {Conference} on {Information} \& {Knowledge} {Management}}, {CIKM} '20, pages 3007--3014, New York, NY, USA. Association for Computing Machinery.
\newblock Event-place: Virtual Event, Ireland.

\bibitem[{Gautam et~al.(2020)Gautam, Mathur, Gosangi, Mahata, Sawhney, and Shah}]{gautamMeTooMAMultiAspectAnnotations2020}
Akash Gautam, Puneet Mathur, Rakesh Gosangi, Debanjan Mahata, Ramit Sawhney, and Rajiv~Ratn Shah. 2020.
\newblock \href {https://ojs.aaai.org/index.php/ICWSM/article/view/7292} {\#{MeTooMA}: {Multi}-{Aspect} {Annotations} of {Tweets} {Related} to the {MeToo} {Movement}}.
\newblock \emph{Proceedings of the International AAAI Conference on Web and Social Media}, 14:209--216.

\bibitem[{Golbeck et~al.(2017)Golbeck, Ashktorab, Banjo, Berlinger, Bhagwan, Buntain, Cheakalos, Geller, Gergory, Gnanasekaran, Gunasekaran, Hoffman, Hottle, Jienjitlert, Khare, Lau, Martindale, Naik, Nixon, Ramachandran, Rogers, Rogers, Sarin, Shahane, Thanki, Vengataraman, Wan, and Wu}]{golbeck_large_2017}
Jennifer Golbeck, Zahra Ashktorab, Rashad~O. Banjo, Alexandra Berlinger, Siddharth Bhagwan, Cody Buntain, Paul Cheakalos, Alicia~A. Geller, Quint Gergory, Rajesh~Kumar Gnanasekaran, Raja~Rajan Gunasekaran, Kelly~M. Hoffman, Jenny Hottle, Vichita Jienjitlert, Shivika Khare, Ryan Lau, Marianna~J. Martindale, Shalmali Naik, Heather~L. Nixon, Piyush Ramachandran, Kristine~M. Rogers, Lisa Rogers, Meghna~Sardana Sarin, Gaurav Shahane, Jayanee Thanki, Priyanka Vengataraman, Zijian Wan, and Derek~Michael Wu. 2017.
\newblock \href {https://doi.org/10.1145/3091478.3091509} {A {Large} {Labeled} {Corpus} for {Online} {Harassment} {Research}}.
\newblock In \emph{Proceedings of the 2017 {ACM} on {Web} {Science} {Conference}}, pages 229--233, Troy New York USA. ACM.

\bibitem[{Grosz and Conde-Cespedes(2020)}]{groszAutomaticDetectionSexist2020}
Dylan Grosz and Patricia Conde-Cespedes. 2020.
\newblock \href {https://doi.org/10.1007/978-3-030-60470-7_11} {Automatic {Detection} of {Sexist} {Statements} {Commonly} {Used} at the {Workplace}}.
\newblock In Wei Lu and Kenny~Q. Zhu, editors, \emph{Trends and {Applications} in {Knowledge} {Discovery} and {Data} {Mining}}, volume 12237, pages 104--115. Springer International Publishing, Cham.

\bibitem[{Hamborg et~al.(2019)Hamborg, Donnay, and Gipp}]{hamborg_automated_2019}
Felix Hamborg, Karsten Donnay, and Bela Gipp. 2019.
\newblock \href {https://doi.org/10.1007/s00799-018-0261-y} {Automated identification of media bias in news articles: an interdisciplinary literature review}.
\newblock \emph{International Journal on Digital Libraries}, 20(4):391--415.

\bibitem[{He et~al.(2019)He, Lee, Ng, and Dahlmeier}]{he2019interactive}
Ruidan He, Wee~Sun Lee, Hwee~Tou Ng, and Daniel Dahlmeier. 2019.
\newblock \href {https://doi.org/10.18653/v1/P19-1048} {An interactive multi-task learning network for end-to-end aspect-based sentiment analysis}.
\newblock In \emph{Proceedings of the 57th Annual Meeting of the Association for Computational Linguistics}, pages 504--515, Florence, Italy. Association for Computational Linguistics.

\bibitem[{He et~al.(2022)He, Mokhberian, and Lerman}]{heInfusingKnowledgeWikipedia2022}
Zihao He, Negar Mokhberian, and Kristina Lerman. 2022.
\newblock \href {https://doi.org/10.48550/ARXIV.2204.03839} {Infusing {Knowledge} from {Wikipedia} to {Enhance} {Stance} {Detection}}.

\bibitem[{Hoel(1971{\natexlab{a}})}]{hoel-71-whole}
Paul~Gerhard Hoel. 1971{\natexlab{a}}.
\newblock \emph{Elementary Statistics}, 3rd edition.
\newblock Wiley series in probability and mathematical statistics. Wiley, New York, Chichester.
\newblock ISBN 0~471~40300.

\bibitem[{Hoel(1971{\natexlab{b}})}]{hoel-71-portion}
Paul~Gerhard Hoel. 1971{\natexlab{b}}.
\newblock \emph{Elementary Statistics}, 3rd edition, Wiley series in probability and mathematical statistics, pages 19--33. Wiley, New York, Chichester.
\newblock ISBN 0~471~40300.

\bibitem[{Hube and Fetahu(2018)}]{hube_detecting_2018}
Christoph Hube and Besnik Fetahu. 2018.
\newblock \href {https://doi.org/10.1145/3184558.3191640} {Detecting {Biased} {Statements} in {Wikipedia}}.
\newblock In \emph{Companion of the {The} {Web} {Conference} 2018 on {The} {Web} {Conference} 2018 - {WWW} '18}, pages 1779--1786, Lyon, France. ACM Press.

\bibitem[{Hube and Fetahu(2019)}]{hube_neural_2019}
Christoph Hube and Besnik Fetahu. 2019.
\newblock \href {https://doi.org/10.1145/3289600.3291018} {Neural {Based} {Statement} {Classification} for {Biased} {Language}}.
\newblock In \emph{Proceedings of the {Twelfth} {ACM} {International} {Conference} on {Web} {Search} and {Data} {Mining}}, {WSDM} '19, pages 195--203, New York, NY, USA. Association for Computing Machinery.
\newblock Event-place: Melbourne VIC, Australia.

\bibitem[{Huguet~Cabot et~al.(2021)Huguet~Cabot, Abadi, Fischer, and Shutova}]{huguet_cabot_us_2021}
Pere-Lluís Huguet~Cabot, David Abadi, Agneta Fischer, and Ekaterina Shutova. 2021.
\newblock \href {https://doi.org/10.18653/v1/2021.eacl-main.165} {Us vs. {Them}: {A} {Dataset} of {Populist} {Attitudes}, {News} {Bias} and {Emotions}}.
\newblock In \emph{Proceedings of the 16th {Conference} of the {European} {Chapter} of the {Association} for {Computational} {Linguistics}: {Main} {Volume}}, pages 1921--1945. Association for Computational Linguistics.
\newblock Event-place: Online.

\bibitem[{Jespersen(1922)}]{Jespersen:1922}
Otto Jespersen. 1922.
\newblock \emph{Language: Its Nature, Development, and Origin}.
\newblock Allen and Unwin.

\bibitem[{Kirstein et~al.(2022)Kirstein, Wahle, Ruas, and Gipp}]{kirstein-etal-2022-analyzing}
Frederic~Thomas Kirstein, Jan~Philip Wahle, Terry Ruas, and Bela Gipp. 2022.
\newblock \href {https://aclanthology.org/2022.gem-1.5} {Analyzing multi-task learning for abstractive text summarization}.
\newblock In \emph{Proceedings of the 2nd Workshop on Natural Language Generation, Evaluation, and Metrics (GEM)}, pages 54--77, Abu Dhabi, United Arab Emirates (Hybrid). Association for Computational Linguistics.

\bibitem[{Kochkina et~al.(2018)Kochkina, Liakata, and Zubiaga}]{kochkinaAllinoneMultitaskLearning2018}
Elena Kochkina, Maria Liakata, and Arkaitz Zubiaga. 2018.
\newblock \href {https://aclanthology.org/C18-1288} {All-in-one: Multi-task learning for rumour verification}.
\newblock In \emph{Proceedings of the 27th International Conference on Computational Linguistics}, pages 3402--3413, Santa Fe, New Mexico, USA. Association for Computational Linguistics.

\bibitem[{Krieger et~al.(2022)Krieger, Spinde, Ruas, Kulshrestha, and Gipp}]{krieger_domain-adaptive_2022}
David Krieger, Timo Spinde, Terry Ruas, Juhi Kulshrestha, and Bela Gipp. 2022.
\newblock \href {https://doi.org/10.1145/3529372.3530932} {A {Domain}-adaptive {Pre}-training {Approach} for {Language} {Bias} {Detection} in {News}}.
\newblock In \emph{2022 {ACM}/{IEEE} {Joint} {Conference} on {Digital} {Libraries} ({JCDL})}, Cologne, Germany.

\bibitem[{Krommyda et~al.(2021{\natexlab{a}})Krommyda, Rigos, Bouklas, and Amditis}]{krommydaExperimentalAnalysisData2021}
Maria Krommyda, Anastasios Rigos, Kostas Bouklas, and Angelos Amditis. 2021{\natexlab{a}}.
\newblock \href {https://doi.org/10.3390/informatics8010019} {An {Experimental} {Analysis} of {Data} {Annotation} {Methodologies} for {Emotion} {Detection} in {Short} {Text} {Posted} on {Social} {Media}}.
\newblock \emph{Informatics}, 8(1):19.

\bibitem[{Krommyda et~al.(2021{\natexlab{b}})Krommyda, Rigos, Bouklas, and Amditis}]{krommyda_experimental_2021}
Maria Krommyda, Anastasios Rigos, Kostas Bouklas, and Angelos Amditis. 2021{\natexlab{b}}.
\newblock \href {https://doi.org/10.3390/informatics8010019} {An {Experimental} {Analysis} of {Data} {Annotation} {Methodologies} for {Emotion} {Detection} in {Short} {Text} {Posted} on {Social} {Media}}.
\newblock \emph{Informatics}, 8(1):19.

\bibitem[{Lee et~al.(2021{\natexlab{a}})Lee, Bang, Madotto, and Fung}]{lee_mitigating_2021}
Nayeon Lee, Yejin Bang, Andrea Madotto, and Pascale Fung. 2021{\natexlab{a}}.
\newblock \href {https://doi.org/10.48550/arXiv.2104.00336} {Mitigating {Media} {Bias} through {Neutral} {Article} {Generation}}.
\newblock \emph{CoRR}, abs/2104.00336.
\newblock \_eprint: 2104.00336.

\bibitem[{Lee et~al.(2021{\natexlab{b}})Lee, Li, Wang, Fung, Ma, Yih, and Khabsa}]{lee-etal-2021-unifying}
Nayeon Lee, Belinda~Z. Li, Sinong Wang, Pascale Fung, Hao Ma, Wen-tau Yih, and Madian Khabsa. 2021{\natexlab{b}}.
\newblock \href {https://doi.org/10.18653/v1/2021.naacl-main.432} {On unifying misinformation detection}.
\newblock In \emph{Proceedings of the 2021 Conference of the North American Chapter of the Association for Computational Linguistics: Human Language Technologies}, pages 5479--5485, Online. Association for Computational Linguistics.

\bibitem[{Lim et~al.(2020)Lim, Jatowt, Färber, and Yoshikawa}]{lim_annotating_2020}
Sora Lim, Adam Jatowt, Michael Färber, and Masatoshi Yoshikawa. 2020.
\newblock \href {https://aclanthology.org/2020.lrec-1.184} {Annotating and {Analyzing} {Biased} {Sentences} in {News} {Articles} using {Crowdsourcing}}.
\newblock In \emph{Proceedings of the {Twelfth} {Language} {Resources} and {Evaluation} {Conference}}, pages 1478--1484, Marseille, France. European Language Resources Association.

\bibitem[{Liu(2012)}]{liuSentimentAnalysisOpinion2012}
Bing Liu. 2012.
\newblock \href {https://doi.org/10.1007/978-3-031-02145-9} {\emph{Sentiment {Analysis} and {Opinion} {Mining}}}.
\newblock Synthesis {Lectures} on {Human} {Language} {Technologies}. Springer International Publishing, Cham.

\bibitem[{Liu et~al.(2021)Liu, Lin, Tan, and Wang}]{liuEnhancingZeroshotFewshot2021}
Rui Liu, Zheng Lin, Yutong Tan, and Weiping Wang. 2021.
\newblock \href {https://doi.org/10.18653/v1/2021.findings-acl.278} {Enhancing {Zero}-shot and {Few}-shot {Stance} {Detection} with {Commonsense} {Knowledge} {Graph}}.
\newblock In \emph{Findings of the {Association} for {Computational} {Linguistics}: {ACL}-{IJCNLP} 2021}, pages 3152--3157, Online. Association for Computational Linguistics.

\bibitem[{Liu et~al.(2019)Liu, Ott, Goyal, Du, Joshi, Chen, Levy, Lewis, Zettlemoyer, and Stoyanov}]{liu_roberta_2019}
Yinhan Liu, Myle Ott, Naman Goyal, Jingfei Du, Mandar Joshi, Danqi Chen, Omer Levy, Mike Lewis, Luke Zettlemoyer, and Veselin Stoyanov. 2019.
\newblock \href {http://arxiv.org/abs/1907.11692} {{RoBERTa}: {A} {Robustly} {Optimized} {BERT} {Pretraining} {Approach}}.
\newblock \emph{CoRR}, abs/1907.11692.
\newblock Eprint: 1907.11692.

\bibitem[{Luo et~al.(2020)Luo, Card, and Jurafsky}]{luoDeSMOGDetectingStance2020}
Yiwei Luo, Dallas Card, and Dan Jurafsky. 2020.
\newblock \href {https://doi.org/10.18653/v1/2020.findings-emnlp.296} {{DeSMOG}: {Detecting} {Stance} in {Media} {On} {Global} {Warming}}.
\newblock In \emph{Findings of the {Association} for {Computational} {Linguistics}: {EMNLP} 2020}, pages 3296--3315, Online. Association for Computational Linguistics.

\bibitem[{Ma et~al.(2021)Ma, Lou, Zhang, Wang, and Vosoughi}]{ma-etal-2021-gradts}
Weicheng Ma, Renze Lou, Kai Zhang, Lili Wang, and Soroush Vosoughi. 2021.
\newblock \href {https://doi.org/10.18653/v1/2021.emnlp-main.455} {{G}rad{TS}: A gradient-based automatic auxiliary task selection method based on transformer networks}.
\newblock In \emph{Proceedings of the 2021 Conference on Empirical Methods in Natural Language Processing}, pages 5621--5632, Online and Punta Cana, Dominican Republic. Association for Computational Linguistics.

\bibitem[{Maas et~al.(2011{\natexlab{a}})Maas, Daly, Pham, Huang, Ng, and Potts}]{maas_learning_2011}
Andrew~L. Maas, Raymond~E. Daly, Peter~T. Pham, Dan Huang, Andrew~Y. Ng, and Christopher Potts. 2011{\natexlab{a}}.
\newblock \href {https://www.aclweb.org/anthology/P11-1015} {Learning {Word} {Vectors} for {Sentiment} {Analysis}}.
\newblock In \emph{Proceedings of the 49th {Annual} {Meeting} of the {Association} for {Computational} {Linguistics}: {Human} {Language} {Technologies}}, pages 142--150, Portland, Oregon, USA. Association for Computational Linguistics.

\bibitem[{Maas et~al.(2011{\natexlab{b}})Maas, Daly, Pham, Huang, Ng, and Potts}]{maasLearningWordVectors2011}
Andrew~L. Maas, Raymond~E. Daly, Peter~T. Pham, Dan Huang, Andrew~Y. Ng, and Christopher Potts. 2011{\natexlab{b}}.
\newblock \href {https://aclanthology.org/P11-1015} {Learning {Word} {Vectors} for {Sentiment} {Analysis}}.
\newblock In \emph{Proceedings of the 49th {Annual} {Meeting} of the {Association} for {Computational} {Linguistics}: {Human} {Language} {Technologies}}, pages 142--150, Portland, Oregon, USA. Association for Computational Linguistics.

\bibitem[{Mathew et~al.(2021)Mathew, Saha, Yimam, Biemann, Goyal, and Mukherjee}]{mathew_hatexplain_2021}
Binny Mathew, Punyajoy Saha, Seid~Muhie Yimam, Chris Biemann, Pawan Goyal, and Animesh Mukherjee. 2021.
\newblock \href {https://doi.org/10.1609/aaai.v35i17.17745} {{HateXplain}: {A} {Benchmark} {Dataset} for {Explainable} {Hate} {Speech} {Detection}}.
\newblock \emph{Proceedings of the AAAI Conference on Artificial Intelligence}, 35(17):14867--14875.

\bibitem[{Miller et~al.(2017)Miller, Feng, Batra, Bordes, Fisch, Lu, Parikh, and Weston}]{millerParlAIDialogResearch2017}
Alexander Miller, Will Feng, Dhruv Batra, Antoine Bordes, Adam Fisch, Jiasen Lu, Devi Parikh, and Jason Weston. 2017.
\newblock \href {https://doi.org/10.18653/v1/D17-2014} {{P}arl{AI}: A dialog research software platform}.
\newblock In \emph{Proceedings of the 2017 Conference on Empirical Methods in Natural Language Processing: System Demonstrations}, pages 79--84, Copenhagen, Denmark. Association for Computational Linguistics.

\bibitem[{Mirzakhmedova et~al.(2023)Mirzakhmedova, Kiesel, Alshomary, Heinrich, Handke, Cai, Valentin, Dastgheib, Ghahroodi, Sadraei, Asgari, Kawaletz, Wachsmuth, and Stein}]{kieseljohannesTouche23HumanValueDetection2022a}
Nailia Mirzakhmedova, Johannes Kiesel, Milad Alshomary, Maximilian Heinrich, Nicolas Handke, Xiaoni Cai, Barriere Valentin, Doratossadat Dastgheib, Omid Ghahroodi, Mohammad~Ali Sadraei, Ehsaneddin Asgari, Lea Kawaletz, Henning Wachsmuth, and Benno Stein. 2023.
\newblock \href {http://arxiv.org/abs/2301.13771} {The touch\'e23-valueeval dataset for identifying human values behind arguments}.

\bibitem[{Mohammad et~al.(2016)Mohammad, Kiritchenko, Sobhani, Zhu, and Cherry}]{mohammad_semeval-2016_2016}
Saif Mohammad, Svetlana Kiritchenko, Parinaz Sobhani, Xiaodan Zhu, and Colin Cherry. 2016.
\newblock \href {https://doi.org/https://doi.org/10.18653/v1/s16-1003} {Semeval-2016 task 6: {Detecting} stance in tweets}.
\newblock In \emph{Proceedings of the 10th {International} {Workshop} on {Semantic} {Evaluation} ({SemEval}-2016)}, pages 31--41.

\bibitem[{Mohammad et~al.(2017)Mohammad, Sobhani, and Kiritchenko}]{mohammadStanceSentimentTweets2017}
Saif~M. Mohammad, Parinaz Sobhani, and Svetlana Kiritchenko. 2017.
\newblock \href {https://doi.org/10.1145/3003433} {Stance and {Sentiment} in {Tweets}}.
\newblock \emph{ACM Transactions on Internet Technology}, 17(3):26:1--26:23.

\bibitem[{Nadeem et~al.(2021{\natexlab{a}})Nadeem, Bethke, and Reddy}]{nadeemStereoSetMeasuringStereotypical2021}
Moin Nadeem, Anna Bethke, and Siva Reddy. 2021{\natexlab{a}}.
\newblock \href {https://doi.org/10.18653/v1/2021.acl-long.416} {{StereoSet}: {Measuring} stereotypical bias in pretrained language models}.
\newblock In \emph{Proceedings of the 59th {Annual} {Meeting} of the {Association} for {Computational} {Linguistics} and the 11th {International} {Joint} {Conference} on {Natural} {Language} {Processing} ({Volume} 1: {Long} {Papers})}, pages 5356--5371, Online. Association for Computational Linguistics.

\bibitem[{Nadeem et~al.(2021{\natexlab{b}})Nadeem, Bethke, and Reddy}]{nadeem_stereoset_2021}
Moin Nadeem, Anna Bethke, and Siva Reddy. 2021{\natexlab{b}}.
\newblock \href {https://doi.org/10.18653/v1/2021.acl-long.416} {{StereoSet}: {Measuring} stereotypical bias in pretrained language models}.
\newblock In \emph{Proceedings of the 59th {Annual} {Meeting} of the {Association} for {Computational} {Linguistics} and the 11th {International} {Joint} {Conference} on {Natural} {Language} {Processing} ({Volume} 1: {Long} {Papers})}, pages 5356--5371. Association for Computational Linguistics.
\newblock Event-place: Online.

\bibitem[{Nangia et~al.(2020)Nangia, Vania, Bhalerao, and Bowman}]{nangiaCrowSPairsChallengeDataset2020}
Nikita Nangia, Clara Vania, Rasika Bhalerao, and Samuel~R. Bowman. 2020.
\newblock \href {https://doi.org/10.18653/v1/2020.emnlp-main.154} {{CrowS}-{Pairs}: {A} {Challenge} {Dataset} for {Measuring} {Social} {Biases} in {Masked} {Language} {Models}}.
\newblock In \emph{Proceedings of the 2020 {Conference} on {Empirical} {Methods} in {Natural} {Language} {Processing} ({EMNLP})}, pages 1953--1967, Online. Association for Computational Linguistics.

\bibitem[{Pang and Lee(2004)}]{pang_sentimental_2004}
Bo~Pang and Lillian Lee. 2004.
\newblock \href {https://doi.org/10.3115/1218955.1218990} {A {Sentimental} {Education}: {Sentiment} {Analysis} {Using} {Subjectivity} {Summarization} {Based} on {Minimum} {Cuts}}.
\newblock In \emph{Proceedings of the 42nd {Annual} {Meeting} on {Association} for {Computational} {Linguistics}}, {ACL} '04, pages 271--es, USA. Association for Computational Linguistics.
\newblock Event-place: Barcelona, Spain.

\bibitem[{Piskorski et~al.(2023)Piskorski, Stefanovitch, Bausier, Faggiani, Linge, Kharazi, Nikolaidis, Teodori, De~Longueville, Doherty, Gonin, Ignat, Kotseva, Mantica, Marcaletti, Rossi, Spadaro, Verile, Da~San~Martino, Alam, and Nakov}]{JRC132862}
Jakub Piskorski, Nicolas Stefanovitch, Valerie-Anne Bausier, Nicolo Faggiani, Jens Linge, Sopho Kharazi, Nikolaos Nikolaidis, Giulia Teodori, Bertrand De~Longueville, Brian Doherty, Jason Gonin, Camelia Ignat, Bonka Kotseva, Eleonora Mantica, Lorena Marcaletti, Enrico Rossi, Alessio Spadaro, Marco Verile, Giovanni Da~San~Martino, Firoj Alam, and Preslav Nakov. 2023.
\newblock \href {https://knowledge4policy.ec.europa.eu/text-mining/news-categorization-framing-persuasion-techniques-annotation-guidelines_en} {News categorization, framing and persuasion techniques: Annotation guidelines}.
\newblock Technical report, European Commission Joint Research Centre, Ispra (Italy).

\bibitem[{Pontiki et~al.(2014)Pontiki, Galanis, Pavlopoulos, Papageorgiou, Androutsopoulos, and Manandhar}]{pontikiSemEval2014TaskAspect2014}
Maria Pontiki, Dimitris Galanis, John Pavlopoulos, Harris Papageorgiou, Ion Androutsopoulos, and Suresh Manandhar. 2014.
\newblock \href {https://doi.org/10.3115/v1/S14-2004} {{SemEval}-2014 {Task} 4: {Aspect} {Based} {Sentiment} {Analysis}}.
\newblock In \emph{Proceedings of the 8th {International} {Workshop} on {Semantic} {Evaluation} ({SemEval} 2014)}, pages 27--35, Dublin, Ireland. Association for Computational Linguistics.

\bibitem[{Pryzant et~al.(2020)Pryzant, Martinez, Dass, Kurohashi, Jurafsky, and Yang}]{pryzant_automatically_2020}
Reid Pryzant, Richard~Diehl Martinez, Nathan Dass, Sadao Kurohashi, Dan Jurafsky, and Diyi Yang. 2020.
\newblock \href {https://doi.org/10.1609/aaai.v34i01.5385} {Automatically neutralizing subjective bias in text}.
\newblock In \emph{Proceedings of the aaai conference on artificial intelligence}, volume~34, pages 480--489.
\newblock Issue: 01.

\bibitem[{Pujari et~al.(2022)Pujari, Oveson, Kulkarni, and Nouri}]{pujariReinforcementGuidedMultiTask2022}
Rajkumar Pujari, Erik Oveson, Priyanka Kulkarni, and Elnaz Nouri. 2022.
\newblock \href {https://ui.adsabs.harvard.edu/abs/2022arXiv220314349P} {Reinforcement {Guided} {Multi}-{Task} {Learning} {Framework} for {Low}-{Resource} {Stereotype} {Detection}}.
\newblock Technical report.
\newblock ADS Bibcode: 2022arXiv220314349P Type: article.

\bibitem[{Puth et~al.(2015)Puth, Neuhäuser, and Ruxton}]{puth_effective_2015}
Marie-Therese Puth, Markus Neuhäuser, and Graeme~D. Ruxton. 2015.
\newblock \href {https://doi.org/10.1016/j.anbehav.2015.01.010} {Effective use of {Spearman}'s and {Kendall}'s correlation coefficients for association between two measured traits}.
\newblock \emph{Animal Behaviour}, 102:77--84.

\bibitem[{Raffel et~al.(2020)Raffel, Shazeer, Roberts, Lee, Narang, Matena, Zhou, Li, and Liu}]{raffel_exploring_2020}
Colin Raffel, Noam Shazeer, Adam Roberts, Katherine Lee, Sharan Narang, Michael Matena, Yanqi Zhou, Wei Li, and Peter~J. Liu. 2020.
\newblock \href {http://jmlr.org/papers/v21/20-074.html} {Exploring the {Limits} of {Transfer} {Learning} with a {Unified} {Text}-to-{Text} {Transformer}}.
\newblock \emph{Journal of Machine Learning Research}, 21(140):1--67.

\bibitem[{Raza et~al.(2022)Raza, Reji, and Ding}]{raza_dbias_2022}
Shaina Raza, Deepak~John Reji, and Chen Ding. 2022.
\newblock \href {https://doi.org/10.1007/s41060-022-00359-4} {Dbias: Detecting biases and ensuring fairness in news articles}.
\newblock \emph{International Journal of Data Science and Analytics}.

\bibitem[{Reardon et~al.(2022)Reardon, Paik, Gao, Parekh, Zhao, Guo, Betke, and Wijaya}]{reardonBUNEmoAffectiveDataset2022}
Carley Reardon, Sejin Paik, Ge~Gao, Meet Parekh, Yanling Zhao, Lei Guo, Margrit Betke, and Derry Wijaya. 2022.
\newblock \href {http://www.lrec-conf.org/proceedings/lrec2022/pdf/2022.lrec-1.267.pdf} {{BU}-{NEmo}: an {Affective} {Dataset} of {Gun} {Violence} {News}}.
\newblock \emph{Proceedings of the 13th Conference on Language Resources and Evaluation (LREC 2022)}, pages 2507--2516.

\bibitem[{Recasens et~al.(2013)Recasens, Danescu-Niculescu-Mizil, and Jurafsky}]{recasens_linguistic_2013}
Marta Recasens, Cristian Danescu-Niculescu-Mizil, and Dan Jurafsky. 2013.
\newblock \href {https://www.aclweb.org/anthology/P13-1162} {Linguistic {Models} for {Analyzing} and {Detecting} {Biased} {Language}}.
\newblock In \emph{Proceedings of the 51st {Annual} {Meeting} of the {Association} for {Computational} {Linguistics} ({Volume} 1: {Long} {Papers})}, volume~1, pages 1650--1659, Sofia, Bulgaria. Association for Computational Linguistics.

\bibitem[{Ruas et~al.(2020)Ruas, Ferreira, Gorsky, França, and Medeiros}]{RuasFGF20}
Terry Ruas, Charles P.~H. Ferreira, William Gorsky, Fabrício~O. França, and Débora M.~R. Medeiros. 2020.
\newblock \href {https://doi.org/10.1016/j.ins.2020.04.048} {Enhanced word embeddings using multi-semantic representation through lexical chains}.
\newblock 532:16--32.

\bibitem[{Ruas et~al.(2019)Ruas, Gorsky, and Aizawa}]{RuasGA19}
Terry Ruas, William Gorsky, and Akiko Aizawa. 2019.
\newblock \href {https://doi.org/10.1016/j.eswa.2019.06.026} {Multi-sense embeddings through a word sense disambiguation process}.
\newblock 136:288--303.

\bibitem[{Ruder and Plank(2017)}]{ruder-plank-2017-learning}
Sebastian Ruder and Barbara Plank. 2017.
\newblock \href {https://doi.org/10.18653/v1/D17-1038} {Learning to select data for transfer learning with {B}ayesian optimization}.
\newblock In \emph{Proceedings of the 2017 Conference on Empirical Methods in Natural Language Processing}, pages 372--382, Copenhagen, Denmark. Association for Computational Linguistics.

\bibitem[{Safi~Samghabadi et~al.(2020)Safi~Samghabadi, Patwa, PYKL, Mukherjee, Das, and Solorio}]{safisamghabadiAggressionMisogynyDetection2020}
Niloofar Safi~Samghabadi, Parth Patwa, Srinivas PYKL, Prerana Mukherjee, Amitava Das, and Thamar Solorio. 2020.
\newblock \href {https://aclanthology.org/2020.trac-1.20} {Aggression and {Misogyny} {Detection} using {BERT}: {A} {Multi}-{Task} {Approach}}.
\newblock In \emph{Proceedings of the {Second} {Workshop} on {Trolling}, {Aggression} and {Cyberbullying}}, pages 126--131, Marseille, France. European Language Resources Association (ELRA).

\bibitem[{Samory et~al.(2020)Samory, Sen, Kohne, Floeck, and Wagner}]{samoryCallMeSexist2020}
Mattia Samory, Indira Sen, Julian Kohne, Fabian Floeck, and Claudia Wagner. 2020.
\newblock \href {https://doi.org/10.48550/ARXIV.2004.12764} {"{Call} me sexist, but...": {Revisiting} {Sexism} {Detection} {Using} {Psychological} {Scales} and {Adversarial} {Samples}}.

\bibitem[{Shu et~al.(2020{\natexlab{a}})Shu, Mahudeswaran, Wang, Lee, and Liu}]{shu_fakenewsnet_2020}
Kai Shu, Deepak Mahudeswaran, Suhang Wang, Dongwon Lee, and Huan Liu. 2020{\natexlab{a}}.
\newblock \href {https://doi.org/10.1089/big.2020.0062} {{FakeNewsNet}: {A} {Data} {Repository} with {News} {Content}, {Social} {Context}, and {Spatiotemporal} {Information} for {Studying} {Fake} {News} on {Social} {Media}}.
\newblock \emph{Big Data}, 8(3):171--188.

\bibitem[{Shu et~al.(2020{\natexlab{b}})Shu, Mahudeswaran, Wang, Lee, and Liu}]{shuFakeNewsNetDataRepository2020}
Kai Shu, Deepak Mahudeswaran, Suhang Wang, Dongwon Lee, and Huan Liu. 2020{\natexlab{b}}.
\newblock \href {https://doi.org/10.1089/big.2020.0062} {{FakeNewsNet}: {A} {Data} {Repository} with {News} {Content}, {Social} {Context}, and {Spatiotemporal} {Information} for {Studying} {Fake} {News} on {Social} {Media}}.
\newblock \emph{Big Data}, 8(3):171--188.

\bibitem[{Singer et~al.(1954--58)Singer, Holmyard, and Hall}]{singer-whole}
Charles~Joseph Singer, E.~J. Holmyard, and A.~R. Hall, editors. 1954--58.
\newblock \emph{A history of technology}.
\newblock Oxford University Press, London.
\newblock 5 vol.

\bibitem[{Sinha and Dasgupta(2021)}]{sinha_determining_2021}
Manjira Sinha and Tirthankar Dasgupta. 2021.
\newblock \href {https://doi.org/10.1145/3459637.3482084} {Determining {Subjective} {Bias} in {Text} through {Linguistically} {Informed} {Transformer} based {Multi}-{Task} {Network}}.
\newblock In \emph{Proceedings of the 30th {ACM} {International} {Conference} on {Information} \& {Knowledge} {Management}}, pages 3418--3422. ACM.
\newblock Event-place: Virtual Event Queensland Australia.

\bibitem[{Sobhani et~al.(2017)Sobhani, Inkpen, and Zhu}]{sobhaniDatasetMultiTargetStance2017}
Parinaz Sobhani, Diana Inkpen, and Xiaodan Zhu. 2017.
\newblock \href {https://aclanthology.org/E17-2088} {A {Dataset} for {Multi}-{Target} {Stance} {Detection}}.
\newblock In \emph{Proceedings of the 15th {Conference} of the {European} {Chapter} of the {Association} for {Computational} {Linguistics}: {Volume} 2, {Short} {Papers}}, pages 551--557, Valencia, Spain. Association for Computational Linguistics.

\bibitem[{Socher et~al.(2013{\natexlab{a}})Socher, Perelygin, Wu, Chuang, Manning, Ng, and Potts}]{socher_recursive_2013}
Richard Socher, Alex Perelygin, Jean Wu, Jason Chuang, Christopher~D. Manning, Andrew Ng, and Christopher Potts. 2013{\natexlab{a}}.
\newblock \href {https://aclanthology.org/D13-1170} {Recursive {Deep} {Models} for {Semantic} {Compositionality} {Over} a {Sentiment} {Treebank}}.
\newblock In \emph{Proceedings of the 2013 {Conference} on {Empirical} {Methods} in {Natural} {Language} {Processing}}, pages 1631--1642, Seattle, Washington, USA. Association for Computational Linguistics.

\bibitem[{Socher et~al.(2013{\natexlab{b}})Socher, Perelygin, Wu, Chuang, Manning, Ng, and Potts}]{socherRecursiveDeepModels2013}
Richard Socher, Alex Perelygin, Jean Wu, Jason Chuang, Christopher~D. Manning, Andrew Ng, and Christopher Potts. 2013{\natexlab{b}}.
\newblock \href {https://aclanthology.org/D13-1170} {Recursive {Deep} {Models} for {Semantic} {Compositionality} {Over} a {Sentiment} {Treebank}}.
\newblock In \emph{Proceedings of the 2013 {Conference} on {Empirical} {Methods} in {Natural} {Language} {Processing}}, pages 1631--1642, Seattle, Washington, USA. Association for Computational Linguistics.

\bibitem[{Spinde et~al.(2023)Spinde, Hinterreiter, Haak, Ruas, Giese, Meuschke, and Gipp}]{spinde2023media}
Timo Spinde, Smilla Hinterreiter, Fabian Haak, Terry Ruas, Helge Giese, Norman Meuschke, and Bela Gipp. 2023.
\newblock The media bias taxonomy: A systematic literature review on the forms and automated detection of media bias.
\newblock \emph{arXiv preprint arXiv:2312.16148}.

\bibitem[{Spinde et~al.(2021{\natexlab{a}})Spinde, Kreuter, Gaissmaier, Hamborg, Gipp, and Giese}]{spinde_you_2021}
Timo Spinde, Christina Kreuter, Wolfgang Gaissmaier, Felix Hamborg, Bela Gipp, and Helge Giese. 2021{\natexlab{a}}.
\newblock \href {https://doi.org/10.1109/JCDL52503.2021.00018} {Do {You} {Think} {It}’s {Biased}? {How} {To} {Ask} {For} {The} {Perception} {Of} {Media} {Bias}}.
\newblock In \emph{Proceedings of the {ACM}/{IEEE} {Joint} {Conference} on {Digital} {Libraries} ({JCDL})}, pages 61--69.

\bibitem[{Spinde et~al.(2021{\natexlab{b}})Spinde, Krieger, Plank, and Gipp}]{spinde_towards_2021}
Timo Spinde, David Krieger, Manu Plank, and Bela Gipp. 2021{\natexlab{b}}.
\newblock \href {https://doi.org/10.1109/JCDL52503.2021.00053} {Towards {A} {Reliable} {Ground}-{Truth} {For} {Biased} {Language} {Detection}}.
\newblock In \emph{Proceedings of the {ACM}/{IEEE}-{CS} {Joint} {Conference} on {Digital} {Libraries} ({JCDL})}, Virtual Event.

\bibitem[{Spinde et~al.(2022)Spinde, Krieger, Ruas, Mitrović, Götz-Hahn, Aizawa, and Gipp}]{SpindeJCDL2022}
Timo Spinde, Jan-David Krieger, Terry Ruas, Jelena Mitrović, Franz Götz-Hahn, Akiko Aizawa, and Bela Gipp. 2022.
\newblock \href {https://doi.org/10.1007/978-3-030-96957-8_20} {Exploiting transformer-based multitask learning for the detection of media bias in news articles}.
\newblock In \emph{Proceedings of the {iConference} 2022}, Virtual event.
\newblock Tex.pubstate: published tex.tppubtype: inproceedings.

\bibitem[{Spinde et~al.(2021{\natexlab{c}})Spinde, Plank, Krieger, Ruas, Gipp, and Aizawa}]{spinde_neural_2021}
Timo Spinde, Manuel Plank, Jan-David Krieger, Terry Ruas, Bela Gipp, and Akiko Aizawa. 2021{\natexlab{c}}.
\newblock \href {https://doi.org/10.18653/v1/2021.findings-emnlp.101} {Neural {Media} {Bias} {Detection} {Using} {Distant} {Supervision} {With} {BABE} - {Bias} {Annotations} {By} {Experts}}.
\newblock In \emph{Findings of the {Association} for {Computational} {Linguistics}: {EMNLP} 2021}, pages 1166--1177, Punta Cana, Dominican Republic. Association for Computational Linguistics.

\bibitem[{Spinde et~al.(2021{\natexlab{d}})Spinde, Rudnitckaia, Mitrović, Hamborg, Granitzer, Gipp, and Donnay}]{spinde_automated_2021}
Timo Spinde, Lada Rudnitckaia, Jelena Mitrović, Felix Hamborg, Michael Granitzer, Bela Gipp, and Karsten Donnay. 2021{\natexlab{d}}.
\newblock \href {https://doi.org/https://doi.org/10.1016/j.ipm.2021.102505} {Automated identification of bias inducing words in news articles using linguistic and context-oriented features}.
\newblock \emph{Information Processing \& Management}, 58(3):102505.

\bibitem[{Sridhar and Getoor(2019)}]{sridharEstimatingCausalEffects2019}
Dhanya Sridhar and Lise Getoor. 2019.
\newblock \href {https://doi.org/10.5555/3367243.3367298} {Estimating causal effects of tone in online debates}.
\newblock In \emph{Proceedings of the 28th International Joint Conference on Artificial Intelligence}, IJCAI'19, page 1872–1878. AAAI Press.

\bibitem[{Strötgen and Gertz(2012)}]{Martin-90}
Jannik Strötgen and Michael Gertz. 2012.
\newblock Temporal tagging on different domains: Challenges, strategies, and gold standards.
\newblock In \emph{Proceedings of the Eight International Conference on Language Resources and Evaluation (LREC'12)}, pages 3746--3753, Istanbul, Turkey. European Language Resource Association (ELRA).

\bibitem[{Superman et~al.(2000)Superman, Batman, Catwoman, and Spiderman}]{Superman-Batman-Catwoman-Spiderman-00}
S.~Superman, B.~Batman, C.~Catwoman, and S.~Spiderman. 2000.
\newblock \emph{Superheroes experiences with books}, 20th edition.
\newblock The Phantom Editors Associates, Gotham City.

\bibitem[{Voigt et~al.(2018{\natexlab{a}})Voigt, Jurgens, Prabhakaran, Jurafsky, and Tsvetkov}]{voigt_rtgender_2018}
Rob Voigt, David Jurgens, Vinodkumar Prabhakaran, Dan Jurafsky, and Yulia Tsvetkov. 2018{\natexlab{a}}.
\newblock \href {https://aclanthology.org/L18-1445} {{RtGender}: {A} {Corpus} for {Studying} {Differential} {Responses} to {Gender}}.
\newblock In \emph{Proceedings of the {Eleventh} {International} {Conference} on {Language} {Resources} and {Evaluation} ({LREC} 2018)}, Miyazaki, Japan. European Language Resources Association (ELRA).

\bibitem[{Voigt et~al.(2018{\natexlab{b}})Voigt, Jurgens, Prabhakaran, Jurafsky, and Tsvetkov}]{voigtRtGenderCorpusStudying2018}
Rob Voigt, David Jurgens, Vinodkumar Prabhakaran, Dan Jurafsky, and Yulia Tsvetkov. 2018{\natexlab{b}}.
\newblock \href {https://aclanthology.org/L18-1445} {{RtGender}: {A} {Corpus} for {Studying} {Differential} {Responses} to {Gender}}.
\newblock In \emph{Proceedings of the {Eleventh} {International} {Conference} on {Language} {Resources} and {Evaluation} ({LREC} 2018)}, Miyazaki, Japan. European Language Resources Association (ELRA).

\bibitem[{Wahle et~al.(2023)Wahle, Gipp, and Ruas}]{WahleGR23}
Jan Wahle, Bela Gipp, and Terry Ruas. 2023.
\newblock \href {https://doi.org/10.18653/v1/2023.emnlp-main.746} {Paraphrase {{Types}} for {{Generation}} and {{Detection}}}.
\newblock In \emph{Proceedings of the 2023 {{Conference}} on {{Empirical Methods}} in {{Natural Language Processing}}}, pages 12148--12164. Association for Computational Linguistics.

\bibitem[{Wahle et~al.(2022)Wahle, Ruas, Kirstein, and Gipp}]{WahleRKG22a}
Jan~Philip Wahle, Terry Ruas, Frederic Kirstein, and Bela Gipp. 2022.
\newblock \href {https://doi.org/10.18653/v1/2022.emnlp-main.62} {How large language models are transforming machine-paraphrase plagiarism}.
\newblock In \emph{Proceedings of the 2022 Conference on Empirical Methods in Natural Language Processing}, pages 952--963, Abu Dhabi, United Arab Emirates. Association for Computational Linguistics.

\bibitem[{Wang et~al.(2019)Wang, Pruksachatkun, Nangia, Singh, Michael, Hill, Levy, and Bowman}]{wang_superglue_2019}
Alex Wang, Yada Pruksachatkun, Nikita Nangia, Amanpreet Singh, Julian Michael, Felix Hill, Omer Levy, and Samuel~R. Bowman. 2019.
\newblock \href {https://proceedings.neurips.cc/paper/2019/hash/4496bf24afe7fab6f046bf4923da8de6-Abstract.html} {{SuperGLUE}: {A} {Stickier} {Benchmark} for {General}-{Purpose} {Language} {Understanding} {Systems}}.
\newblock In \emph{Advances in {Neural} {Information} {Processing} {Systems} 32: {Annual} {Conference} on {Neural} {Information} {Processing} {Systems} 2019, {NeurIPS} 2019, {December} 8-14, 2019, {Vancouver}, {BC}, {Canada}}, pages 3261--3275.

\bibitem[{Wang et~al.(2018)Wang, Singh, Michael, Hill, Levy, and Bowman}]{wang-etal-2018-glue}
Alex Wang, Amanpreet Singh, Julian Michael, Felix Hill, Omer Levy, and Samuel Bowman. 2018.
\newblock \href {https://doi.org/10.18653/v1/W18-5446} {{GLUE}: A multi-task benchmark and analysis platform for natural language understanding}.
\newblock In \emph{Proceedings of the 2018 {EMNLP} Workshop {B}lackbox{NLP}: Analyzing and Interpreting Neural Networks for {NLP}}, pages 353--355, Brussels, Belgium. Association for Computational Linguistics.

\bibitem[{Wang(2017{\natexlab{a}})}]{wang_liar_2017}
William~Yang Wang. 2017{\natexlab{a}}.
\newblock \href {https://doi.org/10.18653/v1/P17-2067} {"{Liar}, {Liar} {Pants} on {Fire}": {A} {New} {Benchmark} {Dataset} for {Fake} {News} {Detection}}.
\newblock In \emph{Proceedings of the 55th {Annual} {Meeting} of the {Association} for {Computational} {Linguistics} ({Volume} 2: {Short} {Papers})}, pages 422--426, Vancouver, Canada. Association for Computational Linguistics.

\bibitem[{Wang(2017{\natexlab{b}})}]{wangLiarLiarPants2017}
William~Yang Wang. 2017{\natexlab{b}}.
\newblock \href {https://doi.org/10.48550/ARXIV.1705.00648} {"{Liar}, {Liar} {Pants} on {Fire}": {A} {New} {Benchmark} {Dataset} for {Fake} {News} {Detection}}.

\bibitem[{Webster et~al.(2018)Webster, Recasens, Axelrod, and Baldridge}]{websterMindGAPBalanced2018}
Kellie Webster, Marta Recasens, Vera Axelrod, and Jason Baldridge. 2018.
\newblock \href {https://doi.org/10.1162/tacl_a_00240} {Mind the {GAP}: {A} {Balanced} {Corpus} of {Gendered} {Ambiguous} {Pronouns}}.
\newblock \emph{Transactions of the Association for Computational Linguistics}, 6:605--617.

\bibitem[{Weinzierl and Harabagiu(2022)}]{weinzierlVaccineLiesNaturalLanguage2022}
Maxwell Weinzierl and Sanda Harabagiu. 2022.
\newblock \href {https://aclanthology.org/2022.lrec-1.753} {{VaccineLies}: {A} {Natural} {Language} {Resource} for {Learning} to {Recognize} {Misinformation} about the {COVID}-19 and {HPV} {Vaccines}}.
\newblock In \emph{Proceedings of the {Thirteenth} {Language} {Resources} and {Evaluation} {Conference}}, pages 6967--6975, Marseille, France. European Language Resources Association.

\bibitem[{Wessel et~al.(2023)Wessel, Horych, Ruas, Aizawa, Gipp, and Spinde}]{WesselMBIB22}
Martin Wessel, Tomas Horych, Terry Ruas, Akiko Aizawa, Bela Gipp, and Timo Spinde. 2023.
\newblock \href {https://doi.org/https://doi.org/10.1145/3539618.3591882} {Introducing {MBIB} - {The} {First} {Media} {Bias} {Identification} {Benchmark} {Task} and {Dataset} {Collection}}.
\newblock In \emph{Proceedings of 46th International ACM SIGIR Conference on Research and Development in Information Retrieval (SIGIR'23)}, New York, NY, USA. ACM.
\newblock ISBN 978-1-4503-9408-6/23/07.

\bibitem[{Wilson(2008)}]{wilsonFinegrainedSubjectivitySentiment2008}
Theresa~Ann Wilson. 2008.
\newblock \href {https://doi.org/10.5555/1559698} {\emph{Fine-Grained Subjectivity and Sentiment Analysis: Recognizing the Intensity, Polarity, and Attitudes of Private States}}.
\newblock Ph.D. thesis, USA.
\newblock AAI3322382.

\bibitem[{Wulczyn et~al.(2017{\natexlab{a}})Wulczyn, Thain, and Dixon}]{wulczyn_ex_2017}
Ellery Wulczyn, Nithum Thain, and Lucas Dixon. 2017{\natexlab{a}}.
\newblock \href {https://doi.org/10.1145/3038912.3052591} {Ex {Machina}: {Personal} {Attacks} {Seen} at {Scale}}.
\newblock In \emph{Proceedings of the 26th {International} {Conference} on {World} {Wide} {Web}, {WWW} 2017, {Perth}, {Australia}, {April} 3-7, 2017}, pages 1391--1399. ACM.

\bibitem[{Wulczyn et~al.(2017{\natexlab{b}})Wulczyn, Thain, and Dixon}]{wulczynExMachinaPersonal2017}
Ellery Wulczyn, Nithum Thain, and Lucas Dixon. 2017{\natexlab{b}}.
\newblock \href {https://doi.org/10.1145/3038912.3052591} {Ex {Machina}: {Personal} {Attacks} {Seen} at {Scale}}.
\newblock In \emph{Proceedings of the 26th {International} {Conference} on {World} {Wide} {Web}}, {WWW} '17, pages 1391--1399, Republic and Canton of Geneva, CHE. International World Wide Web Conferences Steering Committee.

\bibitem[{Yu et~al.(2020)Yu, Kumar, Gupta, Levine, Hausman, and Finn}]{yu2020surgery}
Tianhe Yu, Saurabh Kumar, Abhishek Gupta, Sergey Levine, Karol Hausman, and Chelsea Finn. 2020.
\newblock \href {https://proceedings.neurips.cc/paper/2020/file/3fe78a8acf5fda99de95303940a2420c-Paper.pdf} {Gradient surgery for multi-task learning}.
\newblock In \emph{Advances in Neural Information Processing Systems}, volume~33, pages 5824--5836. Curran Associates, Inc.

\bibitem[{Zhang et~al.(2015{\natexlab{a}})Zhang, Zhao, and LeCun}]{zhangCharacterlevelConvolutionalNetworks2015}
Xiang Zhang, Junbo Zhao, and Yann LeCun. 2015{\natexlab{a}}.
\newblock Character-level convolutional networks for text classification.
\newblock In \emph{Proceedings of the 28th {International} {Conference} on {Neural} {Information} {Processing} {Systems} - {Volume} 1}, {NIPS}'15, pages 649--657, Cambridge, MA, USA. MIT Press.

\bibitem[{Zhang et~al.(2015{\natexlab{b}})Zhang, Zhao, and LeCun}]{zhang_character-level_2015}
Xiang Zhang, Junbo~Jake Zhao, and Yann LeCun. 2015{\natexlab{b}}.
\newblock \href {https://proceedings.neurips.cc/paper/2015/hash/250cf8b51c773f3f8dc8b4be867a9a02-Abstract.html} {Character-level {Convolutional} {Networks} for {Text} {Classification}}.
\newblock In \emph{Advances in {Neural} {Information} {Processing} {Systems} 28: {Annual} {Conference} on {Neural} {Information} {Processing} {Systems} 2015, {December} 7-12, 2015, {Montreal}, {Quebec}, {Canada}}, pages 649--657.

\end{thebibliography}
\appendix
\section{Appendix}\label{sec:app}

\begin{table*}[h]
\tiny

\centering
\begin{adjustbox}{width=\textwidth}
\begin{tabular}{|l|l|l|l|}
\hline
Task Family & Dataset & \# sentences & Task\\
\hline
\hline
\multirow{4}{*}{Subjective bias} & SUBJ \cite{pang_sentimental_2004} & 10.000 & Binary Classification  \\
\cline{2-4}
& Wiki Neutrality Corpus \cite{pryzant_automatically_2020} & 52.036 & Token-level Classification \\
\cline{2-4}
& NewsWCL50 \cite{hamborg_automated_2019} & 731 & Regression  \\
\cline{2-4}
& CW\_HARD \cite{hube_neural_2019} & 6.843 & Binary Classification   \\
\hline
\multirow{7}{*}{News bias} & MultiDimNews \cite{farber_multidimensional_2020} & 2.015 & Multi-Label Classification \\
\cline{2-4}
& \multirow{2}{*}{BASIL \cite{fan_plain_2019}} & \multirow{2}{*}{7.987} & Multi-Class Classification \\
\cline{4-4}
&  &  & Token-Level Classification  \\
\cline{2-4}
& Starbucks \cite{lim_annotating_2020} & 866 & Regression \\
\cline{2-4}
& SemEval2023Task3 \cite{JRC132862} & 5.219 & Binary Classification \\
\cline{2-4}
& \multirow{2}{*}{BABE \cite{spinde_neural_2021}} & \multirow{2}{*}{3.672} & Binary Classification \\
\cline{4-4}
& & & Token-Level Classification \\
\hline
\multirow{12}{*}{Hate speech} & OffensiveLanguage \cite{davidson_automated_2017} & 23.198 & Multi-Class Classification  \\
\cline{2-4}
& OnlineHarassmentDataset \cite{golbeck_large_2017} & 19.613 & Binary Classification\\
\cline{2-4}
& WikiDetoxToxicity \cite{wulczyn_ex_2017} & 138.827 & Regression \\
\cline{2-4}
& \multirow{2}{*}{WikiDetoxAggression \cite{wulczyn_ex_2017}} & \multirow{2}{*}{101.159} & Binary Classification \\
\cline{4-4}
& & & Regression \\
\cline{2-4}
& Jigsaw \cite{jigsaw/conversationaiJigsawUnintendedBias2019} & 101.060 & Binary Classification \\
\cline{2-4}
& MeTooMA \cite{gautamMeTooMAMultiAspectAnnotations2020} & 7.388 & Multi-Label Classification \\
\cline{2-4}
& WikiMadlibs \cite{dixonMeasuringMitigatingUnintended2018} & 74.972 & Binary Classification \\
\cline{2-4}
& \multirow{3}{*}{HateXplain \cite{mathew_hatexplain_2021}} & \multirow{3}{*}{18.962} & Multi-Class Classification \\
\cline{4-4}
& & & Multi-Label Classification \\
\cline{4-4}
& & & Token-Level Classification \\
\cline{2-4}
& HateSpeechTwitter \cite{fountaLargeScaleCrowdsourcing2018} & 48.572 & Multi-Class Classification \\
\hline
\multirow{6}{*}{Gender bias} & GAP \cite{websterMindGAPBalanced2018} & 4.373 & Multi-Class Classification \\
\cline{2-4}
& RtGender \cite{voigt_rtgender_2018} & 21.690 & Binary Classification \\
\cline{2-4}
& MDGender \cite{dinanMultiDimensionalGenderBias2020} & 2.332 & Multi-Class Classification \\
\cline{2-4}
& TRAC2 \cite{safisamghabadiAggressionMisogynyDetection2020} & 3.983 & Binary Classification \\
\cline{2-4}
& Funpedia \cite{millerParlAIDialogResearch2017} & 11.256 & Multi-Class Classification \\
\cline{2-4}
& WizardsOfWikipedia \cite{dinan_wizard_2019} & 29.777 & Multi-Class Classification\\
\hline
\multirow{5}{*}{Sentiment analysis} & SST2 \cite{socher_recursive_2013} & 9.436 & Binary Classification \\
\cline{2-4}
& IMDB \cite{maas_learning_2011} & 13.139 & Binary Classification \\
\cline{2-4}
& MPQA \cite{wilsonFinegrainedSubjectivitySentiment2008} & 3.508 & Binary Classification \\
\cline{2-4}
& SemEval2014 \cite{pontikiSemEval2014TaskAspect2014} & 5.794 & Token-Level Classification \\
\cline{2-4}
& AmazonReviews \cite{zhang_character-level_2015} & 167.396 & Binary Classification \\
\hline
\multirow{4}{*}{Fake news} & LIAR \cite{wang_liar_2017} & 12.742 & Regression \\
\cline{2-4}
& FakeNewsNet \cite{shu_fakenewsnet_2020} & 21.299 & Binary Classification \\
\cline{2-4}
& \multirow{2}{*}{PHEME \cite{kochkinaAllinoneMultitaskLearning2018}} & \multirow{2}{*}{5.022} & Binary Classification \\
\cline{4-4}
& & & Multi-Class Classification \\
\hline
\multirow{5}{*}{Emotionality} & \multirow{2}{*}{GoodNewsEveryone \cite{bostanGoodNewsEveryoneCorpusNews2020}} & \multirow{2}{*}{4.428} & Token-Level Classification \\
\cline{4-4}
& & & Token-Level Classification \\
\cline{2-4}
& BU-NEMO \cite{reardonBUNEmoAffectiveDataset2022} & 12.576 & Multi-Class Classification \\
\cline{2-4}
& EmotionTweets \cite{krommyda_experimental_2021} & 195.744 & Multi-Class Classification \\
\cline{2-4}
& DebateEffects \cite{sridharEstimatingCausalEffects2019} & 6.941 & Regression \\
\hline
\multirow{10}{*}{Group bias} & \multirow{3}{*}{CrowSPairs \cite{nangiaCrowSPairsChallengeDataset2020}} & \multirow{3}{*}{3.009} & Binary Classification \\
\cline{4-4}
& & & Multi-Class Classification \\
\cline{4-4} 
& & & Token-Level Classification \\
\cline{2-4}
& \multirow{2}{*}{StereoSet \cite{nadeem_stereoset_2021}} & \multirow{2}{*}{4.170} & Binary Classification \\
\cline{4-4}
& & & Multi-Class Classification \\
\cline{2-4}
& \multirow{2}{*}{StereotypeDataset \cite{pujariReinforcementGuidedMultiTask2022}} & \multirow{2}{*}{2.208} & Binary Classification \\
\cline{4-4} 
& & & Multi-Label Classification \\
\cline{2-4}
& \multirow{3}{*}{RedditBias \cite{barikeri_redditbias_2021}} & \multirow{3}{*}{10.395} & Binary Classification \\
\cline{4-4}
& & & Multi-Class Classification \\
\cline{4-4}
& & & Token-Level Classification \\
\hline
\multirow{5}{*}{Stance detection} & SemEval2023Task4 \cite{kieseljohannesTouche23HumanValueDetection2022a} & 5.219 & Binary Classification \\
\cline{2-4}
& VaccineLies \cite{weinzierlVaccineLiesNaturalLanguage2022} & 4.497 & Multi-Class Classification \\
\cline{2-4}
& SemEval2016Task6 \cite{mohammad_semeval-2016_2016} & 4.849 & Multi-Class Classification \\
\cline{2-4}
& WTWT \cite{confortiWillTheyWonTTheyVery2020} & 24.681 & Multi-Class Classification \\
\cline{2-4}
& MultiTargetStance \cite{sobhaniDatasetMultiTargetStance2017} & 4.430 & Multi-Class Classification \\
\cline{2-4}
& GWSD \cite{luoDeSMOGDetectingStance2020} & 2.010 & Multi-Class Classification \\
\hline
\hline
& & $\sum$ 1.210.084  & \\
\hline
\end{tabular}
\end{adjustbox}
\caption{References and description to all 59 Tasks (46 datasets) in \lbm collection.}
\label{table:lbm_all}

\end{table*}

\begin{table*}[h]
    \centering

    \begin{tabular}{|c|c|c|c|}
    \hline
    \textbf{MBIB Task} & \textbf{MAGPIE} & \textbf{RoBERTa} & \textbf{ConvBERT} \\
    \hline
    \hline
    Linguistic Bias & \textbf{0.7139} & 0.7076 & 0.7126 \\
    Cognitive Bias & \textbf{0.7086} & 0.7037 & 0.7044 \\
    Text-Level Context Bias & 0.7638 & 0.7646 & 0.7697 \\
    Hatespeech & 0.8747 & 0.8759 & 0.8805 \\
    Gender Bias & \textbf{0.8344} & 0.8322 & 0.8257 \\
    Racial Bias & \textbf{0.8809} & 0.8761 & 0.8772 \\
    Fake News & 0.6709 & 0.6711 & 0.6787 \\
    Political Bias & \textbf{0.7059} & 0.7029 & 0.7041 \\

    \hline
    
    \end{tabular}
        \caption{Performance of MAGPIE and two baselines RoBERTa, and ConvBERT (\sotafull model in \citet{WesselMBIB22}) on the Media Bias Identification Benchmark (MBIB) tasks.}
    \label{table:MBIB_comparison}
\end{table*}

\end{document}